\documentclass[sigconf,nonacm,screen]{acmart}

\usepackage{algorithmicx}
\usepackage{algorithm} 
\usepackage[noend]{algpseudocode} 
\usepackage{amsmath}

\usepackage{graphicx}
\usepackage{subfig}

\usepackage{balance} 
\usepackage{bbding} 
\usepackage{hyperref}
\usepackage{xspace}

\usepackage[capitalize]{cleveref}
\renewcommand{\Cref}[1]{\cref{#1}} 
\renewcommand{\autoref}[1]{\cref{#1}} 
\newcommand{\secref}{$\S\mkern-4mu$}
\crefname{section}{\secref}{$\S\S\mkern-4mu$}
\crefname{figure}{Fig.}{Figs.}
\crefname{table}{Tab.}{Tabs.}
\crefname{appendix}{Appx.}{Apps.}
\crefname{algorithm}{Alg.}{Algs.}

\newcommand{\rp}[1]{{\left(#1\right)}}
\newcommand{\qp}[1]{{\left[#1\right]}}
\newcommand{\cp}[1]{{\left\{#1\right\}}}
\newcommand{\ap}[1]{{\left\langle#1\right\rangle}}

\newcommand{\abs}[1]{{\left|#1\right|}}

\NewDocumentCommand{\such}{}{\text{ s.t. }}
\NewDocumentCommand{\any}{}{\_}

\NewDocumentCommand{\crop}{mm}{\rp{#1\vert_{#2}}}

\newcommand{\bool}{\mathbb{B}}
\newcommand{\true}{\text{\tt True}}
\newcommand{\false}{\text{\tt False}}

\theoremstyle{definition}
    \newtheorem{definition}{Definition}
    \newtheorem{assumption}{Assumption}
    \newtheorem{notation}{Notation}
    
    \newtheorem{observation}{Observation}
\theoremstyle{plain}
    \newtheorem{lemma}{Lemma}
    \newtheorem{theorem}{Theorem}
    \newtheorem{corollary}{Corollary}
    \newtheorem{lresult}{Result}[lemma]

\renewcommand*{\proofname}{Proof}

\newcommand{\event}[3]{
    \ifthenelse
    {\equal{#3}{}}
    {\ap{#1.\textrm{#2}}}
    {\ap{#1.\textrm{#2} \mid #3}}
}

\NewDocumentCommand{\algand}{}{\textbf{ and }}
\NewDocumentCommand{\algst}{}{\textbf{ such that }}
\NewDocumentCommand{\satisfies}{}{\textbf{ satisfies }}
\NewDocumentCommand{\return}{}{\textbf{return }}

\algnewcommand\Instance[2]{\State #1, \textbf{instance} #2}
\algnewcommand\Instances[2]{\State #1, \textbf{multiple instances in} #2}
\algnewcommand\InstanceSystem[3]{\State #1, \textbf{instance} #2, \textbf{system} #3}
\algnewcommand\Trigger[3]{\State \textbf{trigger} $\event{#1}{#2}{#3}$}
\algnewcommand\Schedule[2]{\State \textbf{schedule} $#1$ \textbf{at time} $#2$}

\algsetblockdefx[Implements]{Implements}{EndImplements}{}{}{\textbf{Implements:}}{}
\algsetblockdefx[Uses]{Uses}{EndUses}{}{}{\textbf{Uses:}}{}
\algsetblockdefx[Parameters]{Parameters}{EndParameters}{}{}{\textbf{Parameters:}}{}

\algsetblockdefx[Upon]{Upon}{EndUpon}{}{}[3]{\textbf{upon event} $\event{#1}{#2}{#3}$ \textbf{do}}{}
\algsetblockdefx[UponExists]{UponExists}{EndUponExists}{}{}[2]{\textbf{upon exists} $#1$ \textbf{such that} $#2$ \textbf{do}}{}
\algsetblockdefx[UponCondition]{UponCondition}{EndUponCondition}{}{}[1]{\textbf{upon} $#1$ \textbf{do}}

\algsetblockdefx[Struct]{Struct}{EndStruct}{}{}[1]{\textbf{struct} \texttt{#1} \textbf{contains}}{}
\algsetblockdefx[Procedure]{Procedure}{EndProcedure}{}{}[2]{\textbf{procedure} \emph{#1}($#2$) \textbf{is}}{}

\algdef{SE}[DoUntil]{DoUntil}{EndDoUntil}{\textbf{do}}[1]{\textbf{until} $#1$}
\algsetblockdefx[IfExists]{IfExists}{EndIfExists}{}{}[2]{\textbf{if exists} $#1$ \textbf{such that} $#2$ \textbf{then}}{\textbf{end if}}
\algsetblockdefx[While]{While}{EndWhile}{}{}[1]{\textbf{while} $#1$ \textbf{do}}{\textbf{end while}}
\algsetblockdefx[NTimes]{NTimes}{EndNTimes}{}{}[1]{\textbf{for} $#1$ \textbf{times do}}{\textbf{end for}}
\algsetblockdefx[For]{For}{EndFor}{}{}[3]{\textbf{for} $#1 \in #2..#3$ \textbf{do}}{\textbf{end for}}
\algsetblockdefx[ForAll]{ForAll}{EndForAll}{}{}[2]{\textbf{for all} $#1 \in #2$ \textbf{do}}{\textbf{end for}}

\makeatletter
\ifthenelse{\equal{\ALG@noend}{t}}
	{\algtext*{EndImplements}}
	{}
\ifthenelse{\equal{\ALG@noend}{t}}
	{\algtext*{EndUses}}
	{}
\ifthenelse{\equal{\ALG@noend}{t}}
	{\algtext*{EndParameters}}
	{}
\ifthenelse{\equal{\ALG@noend}{t}}
	{\algtext*{EndUpon}}
	{}
\ifthenelse{\equal{\ALG@noend}{t}}
	{\algtext*{EndUponExists}}
	{}
\ifthenelse{\equal{\ALG@noend}{t}}
	{\algtext*{EndUponCondition}}
	{}
\ifthenelse{\equal{\ALG@noend}{t}}
	{\algtext*{EndStruct}}
	{}
\ifthenelse{\equal{\ALG@noend}{t}}
	{\algtext*{EndProcedure}}
	{}
\ifthenelse{\equal{\ALG@noend}{t}}
	{\algtext*{EndDoUntil}}
	{}
\ifthenelse{\equal{\ALG@noend}{t}}
	{\algtext*{EndFor}}
	{}
\ifthenelse{\equal{\ALG@noend}{t}}
	{\algtext*{EndForAll}}
	{}
\ifthenelse{\equal{\ALG@noend}{t}}
	{\algtext*{EndIf}}
	{}
\makeatother

\NewDocumentCommand{\set}{m}{{\texttt{Set}\rp{#1}}}
\NewDocumentCommand{\map}{mm}{{\texttt{Map}\rp{#1 \rightarrow #2}}}

\NewDocumentCommand{\altime}{}{\mathbb{T}}

\newcommand{\algoblockspacing}{\smallskip}
\AtBeginEnvironment{algorithmic}{\small}

\NewDocumentCommand{\messagedelay}{}{\Delta}
\NewDocumentCommand{\timedrift}{}{\Psi}

\NewDocumentCommand{\servers}{}{\Pi\xspace}
\NewDocumentCommand{\serverssubset}{}{\Sigma\xspace}
\NewDocumentCommand{\serverssubsetsecondary}{}{\Phi\xspace}
\NewDocumentCommand{\server}{}{\pi}
\NewDocumentCommand{\serversecondary}{}{\rho}
\NewDocumentCommand{\servertertiary}{}{\sigma}

\NewDocumentCommand{\clients}{}{\Theta}
\NewDocumentCommand{\client}{}{\theta}

\NewDocumentCommand{\servercount}{}{n}
\NewDocumentCommand{\faultcount}{}{f}

\NewDocumentCommand{\values}{}{\bool}
\NewDocumentCommand{\messages}{}{\mathcal{M}}

\NewDocumentCommand{\con}{}{Representative Binary Consensus\xspace}
\NewDocumentCommand{\conab}{}{RepresentativeBinaryConsensus\xspace}
\NewDocumentCommand{\conal}{}{Blink\xspace}
\NewDocumentCommand{\conin}{}{con\xspace}
\NewDocumentCommand{\condep}{}{dep\xspace}

\NewDocumentCommand{\weakcon}{}{Binary Consensus\xspace}
\NewDocumentCommand{\weakconab}{}{BinaryConsensus\xspace}

\NewDocumentCommand{\validity}{}{Representative Validity\xspace}
\NewDocumentCommand{\weakvalidity}{}{Validity\xspace}

\NewDocumentCommand{\tob}{}{Total Order Broadcast\xspace}
\NewDocumentCommand{\tobab}{}{TotalOrderBroadcast\xspace}
\NewDocumentCommand{\tobal}{}{Flutter\xspace}
\NewDocumentCommand{\tobin}{}{tob\xspace}

\NewDocumentCommand{\betmargin}{}{\epsilon}

\NewDocumentCommand{\universe}{o}{\IfValueTF{#1}
    {{\mathcal{U}\vert_{#1}}}
    {{\mathcal{U}}}}

\NewDocumentCommand{\filtered}{o}{\IfValueTF{#1}
    {{\mathcal{F}\vert_{#1}}}
    {{\mathcal{F}}}}

\NewDocumentCommand{\candidates}{mo}{\IfValueTF{#2}
    {{\mathcal{C}\qp{#1}\vert_{#2}}}
    {{\mathcal{C}\qp{#1}}}}
    
\NewDocumentCommand{\order}{m}{{\mathcal{O}\qp{#1}}}

\NewDocumentCommand{\algo}{}{\mathcal{A}}
\NewDocumentCommand{\exec}{}{\mathcal{E}}

\newcommand{\eg}{e.g.,\xspace}

\newcommand{\ie}{i.e.,\xspace}

\title{Fast Byzantine Total Order Broadcast}

\author{Matteo Monti}
\affiliation{\institution{HES-SO Valais-Wallis \& RAIN}\city{Sierre}\country{Switzerland}}
\orcid{0000-0002-3993-8560}
\email{matteo.monti@hes-so.ch}

\author{Martina Camaioni}
\affiliation{\institution{EPFL}\city{Lausanne}\country{Switzerland}}
\orcid{0009-0000-8106-4732}
\email{martina.camaioni@epfl.ch}

\author{Pierre-Louis Roman}
\affiliation{\institution{Independent}\city{Morges}\country{Switzerland}}
\orcid{0000-0001-5741-1490}
\email{plroman@acm.org}

\date{}

\copyrightyear{2026}
\setcopyright{cc}
\setcctype[4.0]{by}

\thanks{This is the full version of the PODC 2026 paper with DOI \href{https://doi.org/10.1145/3796701.3815946}{10.1145/3796701.3815946}.}

\begin{document}

\begin{abstract}
    This paper presents \tobal, the first Byzantine \tob implementation with a broadcast-to-delivery latency of $2\messagedelay + \betmargin$ time units, $\messagedelay$ being the message delay and $\betmargin$ an arbitrarily small constant margin, when all processes are correct, the network is synchronous, hence local clocks are well-synchronized.
Under the same conditions, state-of-the-art protocols require at least $3\messagedelay$ time units in practical deployments where clients differ from servers.
We prove \tobal's good-case latency is quasi-optimal, meaning it cannot be improved upon by any finite amount.
\tobal is deterministic, leaderless, and signature-free hence quantum-resilient; it assumes partial synchrony and at least $5\faultcount+1$ servers, where $\faultcount$ bounds the number of faults.
Under the hood, \tobal builds upon \conal, a novel \weakcon implementation with \validity, whose fast path enables decisions in $\messagedelay$ time units when all correct servers propose the same value.

\end{abstract}

\maketitle

\section{Introduction}
\label{sec:introduction}

Byzantine Consensus and \tob~\cite{byz-atomic-broadcast-ic95,byz-generals-toplas82} are among the most studied abstractions in the field of distributed computing~\cite{pbft-tocs02,oracles-constantinople-jcrypt05,zyzzyva-sosp07,raynal-async-jacm15,hotstuff-podc19,optimal-fast-bft-podc21,jolteon-ditto-fc22,squad-disc22}, powering State Machine Replication~\cite{lamport_time_cacm78,schneider-smr-csur90} and unlocking reliable computing in the face of arbitrary failures.
Real-world implementations of these abstractions can power a variety of fault-tolerant applications, as exemplified prominently by blockchain-based financial systems~\cite{dfinity-whitepaper-2022,aptos-whitepaper-2022,stellar-consensus-sosp19,espresso-sequencer-whitepaper-2023,sui-whitepaper-2022}.
The two abstractions are closely related~\cite{ct96,dds87}: Consensus has servers agree on a common decision; \tob has servers agree on a totally-ordered sequence of \emph{client-issued messages}.
\tob is often implemented by iterating on Consensus, achieving a common sequence of messages by having servers successively agree on which message(s) to deliver next.

\paragraph{Latency lower bound.}
No implementation of \tob can achieve a broadcast-to-delivery latency shorter than $2\messagedelay$, where $\messagedelay$ bounds the message delay.
To broadcast a message $m$, a client must at least disseminate $m$, taking one message delay.
Once $m$ is disseminated, an additional message delay is required for servers to reach Agreement and Total Order.
A server can deliver $m$ before $2\messagedelay$ only if it gives up on collecting \emph{any} information about $m$ from \emph{any} other server, opening the door for a malicious client to break Agreement simply by disseminating equivocated messages.

Existing blockchain systems separate the set of clients from the set of servers~\cite{dfinity-whitepaper-2022,aptos-whitepaper-2022,sui-whitepaper-2022,espresso-sequencer-whitepaper-2023}, yet, they do not include clients into their latency analysis~\cite{icc-podc22,jolteon-ditto-fc22,shoalpp-nsdi25,bullshark-ccs22,mysticeti-ndss25}.
For instance, the seminal Fast Byzantine (FaB) Paxos implementation~\cite{fab-paxos-dsn05} does not consider clients, rather having servers agree on \emph{server-issued messages} in $2\messagedelay$.
An additional message delay $\messagedelay$ is required for the initial dissemination from clients, leading in practice to a 50\% latency increase.

This paper presents the first Byzantine \tob whose broadcast-to-delivery latency is close to the $2\messagedelay$ lower bound for settings where broadcasting clients differ from delivering servers, as seen in practice.
\tobal is deterministic, signature-free, hence quantum-resilient, assumes partial synchrony and at least $5 \faultcount + 1$ servers, where $\faultcount$ bounds the number of faulty servers.
\tobal uses an \emph{optimistic fast path} and \emph{leader-freedom} to minimize latency.

\paragraph{Optimistic fast path.}
\tobal uses a fast path that activates in the \emph{good case}~\cite{kursawe-optimistic-srds02,jolteon-ditto-fc22,cod-sss22,autobahn-sosp24}.
\tobal upholds all its properties under standard assumptions such as partial synchrony~\cite{dls88} and Byzantine failures.
These assumptions capture well the worst-case behavior of an open-Internet system prone to unpredictable latency spikes and attacks.
In practice, however, these systems are synchronous most of the time, local clocks are well-synchronized, and malicious faults are rare, especially when accountability mechanisms can remove misbehaving processes from the system~\cite{polygraph-icdcs21,bft-forensics-ccs21}.
We refer to these conditions as the \emph{good case}.
These good-case conditions enable \tobal's low latency.

\begin{figure}[tb]
    \centering
    \subfloat[\tobal using \conal.]{
        \includegraphics[height=30mm]{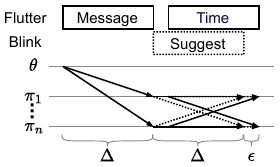}%
        \label{fig:latency-flutter}%
    }
    \hfill
    \subfloat[Fast Byzantine (FaB) Paxos.]{
        \includegraphics[height=30mm]{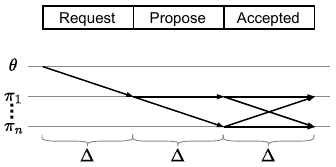}%
        \label{fig:latency-fab-paxos}%
    }
    \caption{%
        \textbf{Good-case latencies of \tobal and FaB Paxos~\cite{fab-paxos-dsn05}.}
        Other fast Byzantine \tob protocols follow FaB Paxos' leader-based pattern: (1) the client $\client$ sends its message to the leader $\server_1$, (2) which disseminates it to all servers $\server_1 \ldots \server_n$, (3) each of which disseminates a confirmation message, (4) upon which all servers decide.
        These protocols have a good-case latency of $3\messagedelay$.
        \tobal skips the forwarding leader, saving a message delay as in Fast Paxos~\cite{lamport-fast-paxos-dc06}, but adds an arbitrarily small margin $\betmargin$ for a good-case latency of $2\messagedelay + \betmargin$.
    }
    \Description[Comparison of the protocol timelines between FaB Paxos and Flutter showing differences in good-case latencies.]{
        FaB Paxos requires three consecutive messages, each taking $\messagedelay$ time unit: Request sent by the client to the leader, Propose sent by the leader to all servers, and Accepted by all servers to all servers.
        The total takes $3 \messagedelay$ time units.
        In comparison, Flutter using Blink also takes three messages but two are mostly overlapping, hence it only takes two consecutive messages and a small added $\batmargin$ time unit in between in the good case: Message sent by the client to all servers at time $t$, Blink's Suggest sent by all servers to all servers at time $t + \messagedelay$, which is overlapping with Time sent at time $t + \messagedelay + \betmargin$ by all servers to all servers.
        The total takes $2 \messagedelay + \betmargin$ time units in the good case.
    }
    \label{fig:latencies}%
\end{figure}

\paragraph{Leader-freedom.}
\tobal is leaderless~\cite{dbft-nca18,leaderless-consensus-jpdc23}, meaning it does not rely on any individual server to make progress.
Leader-freedom brings three key advantages.
First, leaderless algorithms result in lower best-case latency, as they avoid the extra message delay associated with serializing messages through a leader (see \cref{fig:latencies}).
Second, they sidestep the extensive and error-prone logic~\cite{sok-multileader-jsys21} that leader-based solutions require to elect~\cite{stable-leader-election-disc01,pbft-tocs02}, suspect~\cite{polygraph-icdcs21,bft-forensics-ccs21}, and rotate~\cite{banyan-mw24,carousel-fc22} leaders.
Third, they have the potential for improved fairness.
Leader-based implementations are usually vulnerable to a malicious leader reordering messages~\cite{flashboys-sp20,mev-measures-sp22} unless dedicated logic enforces fair ordering~\cite{fair-bft-crypto20,pompe-osdi20,f3b-icdcsw22}.
In \tobal, every message that reaches all correct servers in a timely fashion is delivered, and cannot be delayed.
For comparison, DAG consensus algorithms~\cite{bullshark-ccs22,mysticeti-ndss25,shoalpp-nsdi25,autobahn-sosp24} are also leaderless but focus on providing high-throughput at the cost of higher latency.

\paragraph{\tobal's good-case latency.}
\tobal is the first implementation of Byzantine \tob with a \emph{good-case latency} of $2 \messagedelay + \betmargin$, where $\betmargin$ is an arbitrarily small time interval, \eg the shortest that a processor can represent.
This latency is quasi-optimal; it cannot be improved upon by any finite amount.
In practical deployments where clients differ from servers, \tobal improves latency from state-of-the-art FaB Paxos' $3\messagedelay$ to $2 \messagedelay + \betmargin$ in the good case (see \cref{fig:latencies}).

\tobal bypasses the use of one~\cite{latency-categorization-podc21,bft-smart-dsn14,pbft-tocs02,optimal-fast-bft-podc21,hotstuff-podc19} or several~\cite{fnf-bft-sirocco23,iss-eurosys22,mirbft-jsys22} leaders to reduce latency, rather shifting to clients the responsibility of how messages should be ordered, as done in some crash-stop protocols~\cite{clock-rsm-dsn14,tempo-eurosys21}.
In \tobal, clients tag each broadcast message with a timestamp we call \emph{bet}.
Servers use \emph{\con} to decide \emph{which} messages are accepted: a correct server proposes accepting a message $m$ with bet $b$ if and only if it observes $m$ before $b$; correct servers deliver all accepted messages in increasing order of bet.
\con is a variant of \weakcon whose \validity states that decided values must be proposed by at least $\faultcount + 1$ correct servers (see \cref{subsection:background.abstractions}).
\tobal requires \validity to guarantee Total Order in the presence of slow servers (see \cref{subsection:flutter.mechanisms}).

\tobal uses \conal as \con.
\conal's fast path decides in a single message delay when all correct servers propose the same value.
If not, \conal's slow path delegates decisions to an external \weakcon.
\conal's optimistic latency resembles Bosco's~\cite{bosco-disc08}, a one-step \weakcon in the $7\faultcount + 1$ setting.
\conal's analysis includes a thorough comparison with Bosco and explains why \conal does not infringe on Bosco's result that ``one-step'' $5 \faultcount + 1$ Consensus is impossible (see \cref{section:consensus.bosco}).

\Cref{fig:latencies} depicts how \tobal reaches a good-case latency of $2 \messagedelay + \betmargin$.
First, clients bypass leader(s), directly proposing to servers an ordering for their messages.
A correct client $\client$ broadcasts at time $t$ its message $m$ with a bet $b = t + \tilde \messagedelay + \betmargin$, with $\tilde \messagedelay$ representing $\client$'s best estimate for $\messagedelay$.
Intuitively, $\client$ bets that it can get $m$ to every correct server \emph{strictly before} time $b$, hence the inclusion of $\betmargin$ in the bet.
In good-case conditions where $\client$ correctly estimates the message delay, every correct server receives $\rp{m, b}$ by time $t + \messagedelay < b$.
In such a case, every correct server proposes to accept $m$ to \conal, which activates \conal's fast path and a decision for delivery is reached one message delay later, at time $t + 2\messagedelay$.
By time $t + 2\messagedelay + \betmargin$, every correct server is certain to have observed every message that \conal might decide to accept with a lower bet than $b$, and delivers $m$.

\paragraph{\tobal's worst-case latency.}
\conal's slow path waits for the decision of an underlying instance of \weakcon.
By the Dolev-Strong bound~\cite{ds83}, any deterministic implementation of Consensus has a worst-case latency of at least $O(\faultcount)$.
Importantly, however, \conal is also compatible with any \emph{probabilistic} implementation of \weakcon, such as Ben-Or~\cite{ben-or-83}.
When composed with probabilistic \weakcon, \conal and \tobal inherit its probability-1 liveness, as well as its $O(1)$ expected latency in the worst case.
Intuitively: \conal and \tobal's leaderless design cannot break Dolev-Strong in the deterministic setting, but it sets them up to inherit the constant expected latency of a Ben-Or style algorithm.
To the best of our knowledge, probabilistic \tobal is the first \tob implementation with a worst-case latency that is constant in the number of servers.
For the sake of simplicity, all proofs provided in this paper assume deterministic \con, but they trivially generalize to the probabilistic case.

\paragraph{Roadmap.}
\Cref{sec:background} states the model and defines abstractions.
\Cref{sec:blink} presents \conal, its algorithm, and argues for its correctness, latency, and complexity.
\Cref{sec:flutter} details \tobal's workings, its server algorithm, and argues for its correctness, latency, and complexity.
\Cref{sec:rw} discusses related work and \cref{sec:conclusion} future work.
\Cref{appendix:consensus} fully proves \conal's correctness and good-case latency, and details its differences with Bosco.
\Cref{appendix:tob} fully proves \tobal's correctness, good-case latency and its quasi-optimality, and discusses attacks.

\section{Model and Background}
\label{sec:background}

\subsection{Model}
\label{subsection:background.model}

\paragraph{System, links and scheduling.}
We assume a message-passing system where processes are organized in two sets: \emph{servers} ($\servers$) and \emph{clients} ($\clients$).\footnote{
    Servers and clients can be disjoint or overlap --- the results apply to both settings.
}
We use $\servercount = \abs{\servers}$.
Servers are fixed and every correct process knows $\servers$.
Any two processes communicate via reliable, authenticated, FIFO links, \ie messages are delivered in the order they are sent.
We assume partial synchrony~\cite{dls88}: the message delay between processes is bounded by an unknown constant $\messagedelay$ and the difference between real time and the local time of any process is bounded by an unknown constant $\timedrift$.\footnote{
    We use this definition for convenience.
    It is equivalent to bound the difference between any two local times by $2 \timedrift$, and label as ``real time'' the local time of any correct process.
}
This definition of partial synchrony is stronger than the one stating that $\messagedelay$ is known after an unknown Global Stabilization Time (GST) and thus subsumes it.
We include in $\messagedelay$ the duration of any constant-time computing required to produce a message on the sender side, or deliver a message on the receiver side.
While partially synchronous algorithms rarely distinguish $\messagedelay$ and $\timedrift$, the distinction is relevant considering that $\timedrift \ll \messagedelay$ in practice thanks to accurate clocks, time servers, and synchronization mechanisms using GPS~\cite{gps-clock-ieee99,dtp-sigcomm16,sundial-osdi20,graham-nsdi22}.
The proposed algorithms uphold all their properties even if $\timedrift \geq \messagedelay$.

\paragraph{Adversary and faults.}
Faulty processes are Byzantine~\cite{byz-generals-toplas82}: they may deviate arbitrarily from the algorithm.
Byzantine processes may collude and coordinate their actions.
We assume that at most $\faultcount$ servers are Byzantine, with $n = 5\faultcount + 1$~\cite{qu-hq-sosp05}; we make no assumption about the number of faulty clients.\footnote{
    If client and servers overlap, the client Byzantine threshold may be constrained by the server Byzantine threshold,
    \eg up to one fifth of clients may fail under full overlap.
}
We assume an adaptive adversary that can choose which processes to compromise based on system and protocol state.
The adversary can eavesdrop on any link.
We make no assumption about the adversary's computational power since the proposed algorithms make no use of cryptography.\footnote{
    Authenticated links over the Internet are often implemented with cryptographic key exchange protocols and message authentication codes.
    This may require additional assumptions on the adversary's power to ensure cryptography soundness.
}

\subsection{Abstractions}
\label{subsection:background.abstractions}

\paragraph{\con.}
\emph{\con} is a variant of \weakcon, which has correct servers agree on a binary value $v \in \values$, with $\values = \cp{\true, \false}$.
The interface to a \con instance $\conin$ comprises two events: a \emph{request} noted $\event{\conin}{Propose}{p \in \values}$, which every correct server uses once to submit its proposal $p$; and an \emph{indication} $\event{\conin}{Decide}{d \in \values}$, which a correct server uses to relay that $d$ was decided.
\con ensures \emph{Integrity}: no correct server decides more than once; \emph{Agreement}: no two correct servers decide different values; \emph{Termination}: every correct server eventually decides; \emph{\validity}: the decided value was proposed by at least $\faultcount + 1$ correct servers.
The latter contrasts with \weakcon's Validity stating that a value can be decided if it is proposed by at least one correct server.

\paragraph{\tob.}
\emph{\tob} has correct servers agree on a totally-ordered sequence of client-issued messages.
Let $\messages$ be the set of all possible messages.
The interface to a \tob instance $\tobin$ comprises two events: a \emph{request} noted $\event{\tobin}{Broadcast}{m \in \messages}$, which a correct client uses to broadcast a message, and an \emph{indication} $\event{\tobin}{Deliver}{\client \in \clients, m \in \messages}$, which a correct server uses to deliver $m$ from $\client$.
Correct clients never broadcast the same message twice.
\tob ensures \emph{No Duplication}: no correct server delivers the same message from the same sender twice; \emph{Integrity}: a correct server delivers a message $m$ from a correct client $\client$ only if $\client$ broadcast $m$; \emph{Validity}: all messages broadcast by a correct client are eventually delivered by a correct server; \emph{Agreement and Total Order}: all correct servers eventually deliver the same messages in the same order.

\subsection{Good-Case Conditions}
\label{subsection:background.goodcase}

\paragraph{Good-case assumptions.}
The algorithms presented in this paper uphold all their properties under the assumptions introduced in \cref{subsection:background.model}.
\tobal, however, attains its minimal latency under more stringent conditions we refer to as the \emph{good case}~\cite{sbft-dsn19}.
In the good case: all processes are correct, local computation is instantaneous, and the system is synchronous such that $\messagedelay$ is known and $\timedrift = 0$.
We set $\timedrift = 0$ to streamline the proofs; good-case latency results are generalizable to any synchronous model where $\messagedelay$ and $\timedrift$ are known.

\paragraph{Transient good case.}
All latency proofs in this paper assume that the good case applies to an entire execution of \tobal.
This is done only to streamline the analysis.
In practice, a real-world system might oscillate between extended periods where the good case assumptions apply, and brief intervals where they do not.
During the latter, servers might disagree on which messages are timely and should be accepted.
This forces the processing of those messages out of \conal's fast path.
Once the system returns to the good-case state, its pipeline might still contain some of those slow-path messages.
As \con always guarantees Termination, however, those messages are eventually processed.
As soon as that happens, \conal resumes delivering messages with minimal latency.

\section{Blink}
\label{sec:blink}

This section presents the proposed \con implementation \conal.
\conal's fast path is central to the design of \tobal (see \cref{sec:flutter}) whose good-case execution relies on all correct servers proposing the same value in \conal.

\Cref{section:consensus.overview} overviews \conal's workings and \cref{section:consensus.algorithm} presents its simple algorithm.
\Cref{section:consensus.correctness.sketches} argues about \conal's correctness, \cref{section:consensus.latency.sketches} about its latency, and \cref{section:consensus.complexity} about its complexity.
\Cref{appendix:consensus} further contains full proofs of \conal's correctness and latency, and a detailed discussion of the differences with Bosco~\cite{bosco-disc08}.

\begin{figure}[t]
    \centering
    \includegraphics[height=28mm]{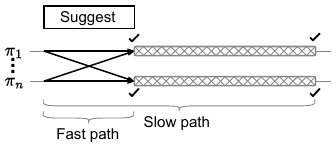}%
    \caption{%
        \textbf{\conal algorithm.}
        A correct server disseminates its proposal as a suggestion, then proposes to \weakcon whichever value it was suggested the most.
        A correct server decides what \weakcon decides.
        The fast path activates when a correct server receives $4 \faultcount + 1$ matching suggestions for a value; the server decides that value.
        Notation:
        a correct server's engagement in \weakcon is marked by a crossed bar; proposals and decisions are indicated at the beginning and end of each bar; \Checkmark represents $\true$.
    }
    \Description[Timeline of Blink.]{
        Blink only uses a single all-to-all message in its fast path, Suggest, and can thus terminate in $\messagedelay$ time unit.
        If the fast path conditions are not met, the slow path is activated.
        In the slow path, an external Binary Consensus is instantiated and Blink decides whatever that instance eventually decides.
    }
    \label{fig:blink-protocol}%
\end{figure}

\subsection{Overview}
\label{section:consensus.overview}

\paragraph{In a blink.}
\conal builds \con on top of a \weakcon instance $\condep$, while enabling a single-message-delay fast path when all correct servers propose the same value.
\conal does so via a single all-to-all communication, wherein each correct server disseminates its proposal as a \emph{suggestion} (see \cref{fig:blink-protocol}).
Upon collecting $4 \faultcount + 1$ suggestions, a correct server proposes to $\condep$ the majority value, \ie the value suggested at least $2\faultcount + 1$ times.
A correct server always decides whatever $\condep$ decides.
Fast path: upon collecting $4 \faultcount + 1$ \emph{matching} suggestions (\ie $4 \faultcount + 1$ \texttt{True} or $4 \faultcount + 1$ \texttt{False}), a correct server instantly decides that value.

\paragraph{\validity.}
All-to-all suggestions are the key to upgrade $\condep$'s \weakvalidity into \validity.
\conal decides a value $v$ only if $\condep$ decides $v$.
By the \weakvalidity of $\condep$, if $\condep$ decides $v$, then at least one correct server proposed $v$ to $\condep$.
In turn, that correct server must have received at least $2 \faultcount + 1$ suggestions for $v$, meaning $v$ was proposed by at least $\faultcount + 1$ correct servers.

\paragraph{Fast path.}
Suggestions also unlock \conal's fast path.
Even after submitting a value to $\condep$, a correct server keeps gathering suggestions; if it ever collects $4\faultcount + 1$ suggestions for $v$, it immediately decides $v$ without waiting for $\condep$'s slower output.
No need to wait: out of those $4\faultcount + 1$ suggestions, at least $3\faultcount + 1$ came from correct servers, hence no more than $f$ correct servers proposed the opposite value $\bar v$. 
By \validity, $\bar v$ cannot be decided.
If every correct server proposes the same value $v$, every correct server collects $4\faultcount + 1$ suggestions for $v$ and decides in one message delay.

\paragraph{Slow path.}
The slow path of \conal's inherits $\condep$'s latency: if proposals do not match, a correct server might never collect $4 \faultcount + 1$ matching suggestions, thus being forced to wait for $\condep$'s decision.
If $\condep$ is deterministic, its latency is at least $O(\faultcount)$~\cite{ds83}.
Importantly, however, \conal is fully compatible with a probabilistic implementation of $\condep$, such as Ben-Or~\cite{ben-or-83}.
When composed with a probabilistic $\condep$~\cite{ben-or-83,crain-two-more-o1-arxiv,raynal-async-jacm15}, \conal inherits its probability-1 liveness, but the expected latency of its slow path is reduced to $O(1)$.
This importantly impacts the worst-case latency of \tobal (see \cref{subsection:tob.latencysketches}).
For the sake of simplicity, all proofs in this paper assume a deterministic $\condep$, but they trivially generalize to the probabilistic case.

\subsection{Algorithm}
\label{section:consensus.algorithm}

\conal implements a \con instance $\conin$ on top of a \weakcon instance $\condep$, as described in \cref{algorithm:conal}.
Upon proposing a value $v$ (line \ref{line:conal.propose}), a correct server $\server$ sends $v$ to all servers by means of a $\texttt{Suggest}$ message (lines \ref{line:conal.suggestfor} and \ref{line:conal.sendsuggest}); $\server$ records the suggestions it receives (line \ref{line:conal.deliversuggest}) in its $suggestions$ map (line \ref{line:conal.recordsuggestion}).
Upon first collecting $4 \faultcount + 1$ suggestions (line \ref{line:conal.depproposecondition}), $\server$ identifies whichever \emph{majority value} $m$ it was suggested the most (line \ref{line:conal.suggestionmajority}), then proposes $m$ to $\condep$ (line \ref{line:conal.deppropose}).
Note that, as $4 \faultcount + 1$ is odd and only two values are possible, $\true$ and $\false$, one and only one value is guaranteed to be backed by at least $2\faultcount + 1$ suggestions.
$\server$ decides $v$ (line \ref{line:conal.decidefast} or \ref{line:conal.decideslow}) upon collecting $4\faultcount + 1$ suggestions for $v$ (fast path, line \ref{line:conal.fastcondition}) or upon $\condep$ deciding $v$ (slow path, line \ref{line:conal.depdecide}).

\begin{algorithm}[t] 
\begin{algorithmic}[1]
\Implements
    \Instance{\conab}{\conin}
\EndImplements
\algoblockspacing

\Uses
    \Instance{AuthenticatedLinks}{al}
    \Instance{\weakconab}{\condep}
\EndUses
\algoblockspacing

\Upon{\conin}{Init}{}
    \State $suggestions: \map{\servers}{\values} = \cp{}$ \label{line:conal.initsuggestions}
    \State $proposed: \bool = \false$ \label{line:conal.initproposed}
    \State $decided: \bool = \false$ \label{line:conal.initdecided}
\EndUpon
\algoblockspacing

\Upon{\conin}{Propose}{v} \label{line:conal.propose}
    \ForAll{\server}{\servers} \label{line:conal.suggestfor}
        \Trigger{al}{Send}{\server, \qp{\texttt{Suggest}, v}} \label{line:conal.sendsuggest}
    \EndForAll
\EndUpon
\algoblockspacing

\Upon{al}{Deliver}{\servertertiary, \qp{\texttt{Suggest}, v}} \label{line:conal.deliversuggest}
    \State $suggestions\qp{\servertertiary} = v$ \label{line:conal.recordsuggestion}
\EndUpon
\algoblockspacing

\UponCondition{\abs{suggestions} \geq 4\faultcount + 1 \algand proposed = \false} \label{line:conal.depproposecondition}
    \State $m = v \algst \abs{\cp{\server \such suggestions\qp{\server} = v}} \geq 2\faultcount + 1$ \label{line:conal.suggestionmajority}
    \Trigger{\condep}{Propose}{m} \label{line:conal.deppropose}
    \State $proposed = \true$ \label{line:conal.setproposed}
\EndUponCondition
\algoblockspacing

\UponExists{v}{\abs{\cp{\server \such suggestions\qp{\server} = v}} \geq 4 \faultcount + 1} \label{line:conal.fastcondition}
    \If{$decided = \false$} \label{line:conal.checkdecidedfast}
        \Trigger{\conin}{Decide}{v} \label{line:conal.decidefast}
        \State $decided = \true$ \label{line:conal.setdecidedfast}
    \EndIf
\EndUponExists
\algoblockspacing

\Upon{\condep}{Decide}{v} \label{line:conal.depdecide}
    \If{$decided = \false$} \label{line:conal.checkdecidedslow}
        \Trigger{\conin}{Decide}{v} \label{line:conal.decideslow}
        \State $decided = \true$ \label{line:conal.setdecidedslow}
    \EndIf
\EndUpon
\end{algorithmic}
\caption{\conal}
\label{algorithm:conal}
\end{algorithm}

\subsection{Correctness Sketches}
\label{section:consensus.correctness.sketches}

\Cref{section:consensus.correctness} proves correctness to the fullest extent of formal detail.

\paragraph{Termination.}
As we assume at least $4\faultcount + 1$ correct servers, every correct server is guaranteed to eventually collect $4\faultcount + 1$ suggestions and submit a majority value to $\condep$.
By $\condep$'s Termination, every correct server is sure to eventually trigger $\condep$'s decision and decide.

\paragraph{\validity.}
If less than $\faultcount + 1$ correct servers propose some value $v$, no correct server can collect the $2\faultcount + 1$ suggestions required to propose $v$ to $\condep$, or the $4\faultcount + 1$ required to fast decide $v$; by $\condep$'s \weakvalidity, no correct server slow decides $v$ either.

\paragraph{Integrity.}
Decisions are guarded by a $decided$ flag, set when a correct server first decides.

\paragraph{Agreement.}
A correct server decides $v$ via the fast path only if it collected at least $4\faultcount + 1$ suggestions for $v$.
In that case, no more than $\faultcount$ correct servers proposed $\bar v$, and no correct server ever collects $2\faultcount + 1$ suggestions for $\bar v$.
As a result, no correct server decides $\bar v$ via the fast path or proposes $\bar v$ to $\condep$.
Because $\condep$ satisfies \weakvalidity, $\condep$ cannot decide $\bar v$.
In summary, if any correct server decides $v$ via the fast path, no correct server decides $\bar v$.
If all correct server decide via the slow path, Agreement follows immediately from the Agreement of $\condep$.

\subsection{Latency Sketches}
\label{section:consensus.latency.sketches}

\paragraph{Fast path latency.}
We fully prove in \cref{subsection:consensus.latency} that, if every correct server proposes the same value $v$ to \conal, then every correct server decides in one message delay.
The result is intuitive: as we assume at least $4\faultcount + 1$ correct servers, every correct server delivers $4\faultcount + 1$ suggestions for $v$ in one message delay and immediately decides $v$.

\paragraph{Slow path latency.}
If \conal fails to decide via the fast path, \conal's latency is one message delay higher than its underlying implementation $\condep$ of \weakcon.
We conjecture this being the unavoidable cost of upgrading \weakvalidity to \validity.

\subsection{Complexity}
\label{section:consensus.complexity}

\conal involves a single step of constant-sized, all-to-all communication, resulting in $O(\servercount^2)$ bit complexity and matching the complexity lower bound for \weakcon~\cite{dr85}.
Meaning, \conal never bottlenecks \weakcon\xspace --- rather, \conal inherits the complexity of the \weakcon implementation it depends upon.

\section{Flutter}
\label{sec:flutter}

This section describes the proposed \tob implementation \tobal.
\tobal relies on \con (see \cref{subsection:background.abstractions}) to decide which messages are delivered.
For \con, \tobal uses \conal (see \cref{sec:blink}) whose fast path enables \tobal's good-case latency of $2 \messagedelay + \betmargin$.

\Cref{subsection:flutter.overview} presents the intuition behind \tobal and
\cref{subsection:flutter.mechanisms} its inner mechanisms.
\Cref{subsection:flutter.timeline} depicts an example execution as viewed by a correct server.
\Cref{subsection:flutter.algorithm} details \tobal's server algorithm.
\Cref{subsection:tob.correctnesssketches} proposes proof sketches arguing for correctness.
\Cref{subsection:tob.latencysketches} briefly argues about latency and \cref{section:flutter.quasioptimal.sketches} about its quasi-optimality.
\Cref{subsection:tob.complexity} discusses complexity.
\Cref{appendix:tob} further presents \tobal's client algorithm, and details full proofs of correctness, of its good-case latency and of its quasi-optimality, as well as its reaction to denial of services.

\begin{figure*}[t]
    \centering
    \includegraphics[height=28mm]{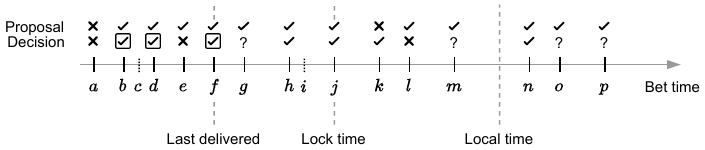}%
    \caption{%
        \textbf{Example client messages in \tobal, noted $a$ to $p$, as seen by a correct server at a given time.}
        Messages are organized on a timeline, appearing at the time of their bet.
        The Consensus status of each message is represented by the two symbols on top of the message: the server's proposal on top (\Checkmark to deliver the message, \textsf{X} to reject it), \con's decision on the bottom (\Checkmark or \textsf{X} to deliver or reject, \textbf{?} if \con is still running).
        The server already delivered all messages whose decision is boxed.
        Messages marked by a solid line are candidates, those marked by a dashed line are not.
    }
    \Description[Timeline of a Flutter execution on a correct server.]{
        The figure shows a series of messages ordered by bet as observed by a correct server.
        The timeline is split by three delimiters: the time of the last delivered message, the lock time, and the local time.
        Every message that is a candidate is tagged by the value that the correct server proposed to consensus, and the value decided by the corresponding consensus instance.
        A proposal can be $\true$ or $\false$ for a message, while a decision can $\true$, $\false$, or pending.
        Non-candidate messages have no such values associated to them.
        The goal of the figure is to show that not all messages are candidates, that different messages may have different proposal and decision values associated to them, and how the delimiters impact these values.
    }
    \label{fig:flutter-server-timeline}%
\end{figure*}

\subsection{Overview}
\label{subsection:flutter.overview}

\tob implementations usually delegate the responsibility of ordering messages to one~\cite{bft-smart-dsn14,pbft-tocs02,fab-paxos-dsn05,hotstuff-podc19} or several~\cite{caesar-dsn17,clock-rsm-dsn14,tempo-eurosys21,mencius-osdi08,egalitarian-paxos-sosp13,iss-eurosys22} leaders.
By contrast, \tobal is leaderless and focuses on providing low latency.
This section outlines \tobal's approach to leader-freedom using timestamps, and underlines the \emph{problem of convergence} that such an approach entails.
\Cref{subsection:flutter.mechanisms} discusses how \tobal solves this problem of convergence.

\paragraph{Bets.}
To bypass leaders, \tobal shifts to clients the responsibility to propose how messages should be ordered~\cite{clock-rsm-dsn14,tempo-eurosys21}.
Correct clients attach to each message a timestamp, which we call \emph{bet}.
Correct servers always deliver messages in increasing order of bet.
By issuing a message $m$ with bet $b$, a client proposes that $m$ should be delivered \emph{with rank} $b$, \ie that $b$ should determine where $m$ appears in the common sequence of delivered messages.

\paragraph{Algorithm intuition.}
Clients are free to propose the delivery of any message with any bet; correct servers use \con to determine \emph{which} proposals are accepted.
The algorithm operates as follows: upon broadcasting a message $m$, a correct client $\client$ checks its local clock, estimates some future bet time $b$ by which it expects to reach all correct servers, then disseminates $\rp{m, b}$.
To reach all correct servers in time, $\client$ sets $b = t + \tilde \messagedelay + \betmargin$, where $t$ is $\client$'s local time, $\tilde \messagedelay$ is $\client$'s best estimate of the message delay $\messagedelay$, and $\betmargin$ is an arbitrarily small constant margin.
While $\client$ cannot reliably determine $\messagedelay$ in partial synchrony, it can in practice obtain $\tilde \messagedelay$ heuristically in the good case (see \cref{subsection:background.goodcase}), \eg by measuring its latency to $4\faultcount + 1$ servers.
The margin $\betmargin$ is added to the bet to ensure that when a correct server $\server$ receives $m$, $b$ is \emph{strictly} in the future according to $\server$'s local clock.
The margin must be strictly positive but can be arbitrarily small; in practice, $\betmargin$ would be the smallest unit of time that $\client$ can represent.
Should $\rp{m, b}$ be rejected for being too late (see below), $\client$ updates $b$ to a more conservative value and tries again.
Because $\messagedelay$ is fixed, $\client$ is guaranteed to eventually reach all correct servers in time and get $m$ delivered.

Upon receiving $\rp{m, b}$, correct servers use \con to decide whether to deliver $m$ with rank $b$.
A correct server $\server$ proposes $\true$ to \con if and only if $\server$ receives $\rp{m, b}$ strictly before $b$ according to $\server$'s local clock; $\server$ delivers $m$ with rank $b$ if and only if \con decides $\true$ for $\rp{m, b}$.
$\rp{m, b}$ may be rejected by \con if $\client$ mis-bets or crashes mid-dissemination (see \cref{fig:flutter-protocol-bad}).

\paragraph{The problem of convergence.}
\tob has correct servers deliver (1) the same messages (Agreement), (2) in the same order (Total Order).
In most implementations, these two conditions are facilitated by a leader that tags each message with a monotonic counter.
This counter enables correct servers to (1) detect missing messages by finding gaps in the enumeration, and (2) achieve a common message order.
Bets ensure condition (2) but not (1): unlike \emph{contiguous} counters, bets are \emph{sparse} in nature.
A server cannot determine to have missed a message from bets alone.

In \tobal, a correct server must (1) deliver all messages for which \con decides $\true$, (2) in increasing order of bet.
To uphold these guarantees, a correct server $\server$ that delivers a message $m$ with rank $b$ must be certain to have already delivered every message predating $m$, \ie every message whose bet is earlier than $b$, for which \con may decide $\true$.
Correct servers must \emph{converge}.
For convergence, it is not sufficient for $\server$ to wait for \con to decide on all the messages that $\server$ \emph{already} observed predating $m$; $\server$ must be sure that \con would reject any new message $m'$ predating $m$ that $\server$ could discover \emph{in the future}.
Failing that, $\server$ could later find some $m'$ it should have delivered before $m$, forcing $\server$ to skip $m'$, breaking Agreement, or deliver $m'$ out of order, breaking Total Order.

Bets set by clients include an extra $\betmargin$ precisely because of this problem.
Without $\betmargin$, it would be possible for different correct servers to receive a message with bet $b$ concurrently to the delivery of other messages with equal bets $b$, which may lead to correct servers ordering messages with the same bet differently, thus breaking Total Order.
With $\betmargin$, correct servers are guaranteed to have received all deliverable messages before they are ordered for delivery,

Messages with the same bet must further be ordered among themselves, \eg by lexicographic order of client identifiers and message payloads (see \cref{assumption:tob.ordering} in \cref{subsection:tob.agreementtotalorder}).

\subsection{Achieving Convergence in Flutter}
\label{subsection:flutter.mechanisms}

\tobal solves the problem of convergence by having servers update each other on the messages and bets that they \emph{observe}, as well as their \emph{local time}.
At a glance: by the \validity of \con, if enough servers announce having reached some time $t$ without ever observing some $(m, b)$, such that $b < t$, then $(m, b)$ is rejected.
This guarantee allows a correct server to converge in finite time on a finite set of \emph{candidates}, which includes every message predating $t$ that \con could ever accept delivering.
Servers deliver, in increasing order of bet, all converged candidates that \con decides to accept.

\paragraph{Observations.}
Upon first receiving a message $m$ with bet $b$ from any source (server or client), a correct server relays $(m, b)$ to every other server via an $\texttt{Observe}$ message.

\paragraph{Local time and lock time.}
Regularly, a correct server $\server$ \emph{announces its local time} to other servers via a $\texttt{Time}$ message.
At any point in time, we call $\server$'s \emph{lock time} the highest local time that $\server$ witnessed being reached by at least $4\faultcount + 1$ servers.
As $4\faultcount + 1$ correct servers regularly disseminate their growing local time, $\server$'s lock time is guaranteed to grow indefinitely.

\paragraph{The role of \validity.}
As discussed in \cref{subsection:flutter.overview}, a correct server $\server$ proposes accepting some $\rp{m, b}$ only if it receives $\rp{m, b}$ before time $b$ according to its local clock.
Hence, once $\server$ announces its local time being $t$, $\server$ is guaranteed to propose rejecting any $\rp{m', b' < t}$ that it did not already receive, propose accepting, and relay.
\tobal solves convergence by generalizing this observation to a quorum of servers.
If at least $4 \faultcount + 1$ servers announce to $\server$ that they reached some time $t$ without \emph{any} of them relaying $\rp{m', b' < t}$, then $\rp{m', b'}$ will certainly be rejected.
Indeed, out of this $4 \faultcount + 1$ quorum, there are at least $3 \faultcount + 1$ correct servers that reached local time $t$ without ever observing $\rp{m', b'}$, and will propose rejecting it.
By the \validity of \con, $\rp{m', b'}$ will be rejected.
This is why \validity matters; a weaker \weakcon could accept $\rp{m', b'}$ due to a single slow-but-correct server.

\paragraph{Candidates.}
Putting all of the above together, a correct server $\server$ maintains a set of \emph{candidates} containing every $\rp{m, b}$ that $\server$ observed before its lock time reached $b$.
By lock time's definition, if some $\rp{m', b'}$ is not among $\server$'s candidates, then at least $4 \faultcount + 1$ servers announced reaching time $b'$ without ever observing and relaying $\rp{m', b'}$, and $m'$ will be rejected.
Once $\server$'s lock time reaches $t$, the set of candidates predating $t$ is fixed: $\server$ waits for \con to decide which candidates are accepted, and delivers each message in increasing order of bet.

\subsection{A Server's Timeline}
\label{subsection:flutter.timeline}

\Cref{fig:flutter-server-timeline} depicts an example timeline of the client-issued messages a correct server $\server$ has discovered at a given time.
Each message appears at the time of its bet; the same message would appear at the same location on the timeline of any other correct server.

The timeline is split by three dividers: $\server$'s local time, $\server$'s lock time, and the bet of the last message delivered by $\server$.
Assuming well-synchronized server clocks where $\timedrift \ll \messagedelay$, $\server$'s local time should be greater than its lock time by approximately one message delay.
If every correct server disseminates its local time at time $t$, $\server$ can expect to receive all updates one message delay later, updating its lock time to $t$ as its local time reaches approximately $t + \messagedelay$.
Clock drift and quicker-than-expected \texttt{Time} updates can perturb this order --- indeed, $\server$'s lock time can overtake its local time if $\server$'s local clock experiences a large enough delay.
The following applies regardless of the order between $\server$'s local and lock time.

The local time delimits which messages $\server$ proposes to accept.
Upon first receiving a message $m$ with bet $b$, $\server$ proposes to accept $m$, by submitting $\true$ to \con, if and only if $b$ is greater than $\server$'s local time.
Note that every message right of $\server$'s local time has a positive proposal, as they have all been observed before $\server$'s local time reached their bet.
This does not apply to all messages on the left of $\server$'s local time.
In the example, messages $a$ and $k$ have been received too late by $\server$.
Despite $\server$'s negative proposal, \con still decided to accept $k$ because at least $\faultcount + 1$ correct servers observed $k$ in time.

The lock time delimits the messages that $\server$ marks as candidates.
Upon first receiving a message $m$ with bet $b$, $\server$ marks $m$ as a candidate if and only if $b$ is greater than $\server$'s lock time.
Every message right of $\server$'s lock time is a candidate, as they have all been observed before $\server$'s lock time reached their bet.
Non-candidate messages are guaranteed to be rejected by Consensus, \eg $c$ and $i$ in the example.

The lock time also delimits which messages $\server$ can deliver.
As no new candidates can ever appear with a bet smaller than $\server$'s lock time, $\server$ can deliver, in increasing order of bet, all the candidates left of its lock time that \con decided to accept.
In the example, the last message delivered by $\server$ is $f$.
The next observed message $g$ is a candidate that has no Consensus decision associated to it yet.
Because $\server$ has to wait for $g$'s decision before moving on to subsequent messages, $\server$ cannot yet deliver $h$ despite \con having decided to accept it.
By the Termination of \con, and because no new candidates can be added with a bet smaller than $\server$'s lock time, $\server$ is guaranteed to sequentially progress through each candidate, eventually delivering every message that \con decides should be delivered, in increasing order of bet, \eg as already done with messages $b$, $d$ and $f$.

\subsection{Server Algorithm}
\label{subsection:flutter.algorithm}

\tobal implements \tob using multiple instances $\conin$ of \con.
\Cref{algorithm:tobalserver} describes \tobal's server side
and \cref{algorithm:tobalclient} in \cref{subsection:tob.client.algorithm} its simple client side.

\paragraph{Variables.}
A correct server $\server$ maintains: an $observed$ set (line \ref{line:tob.initobserved}), containing all $(client, message, bet)$ tuples $\server$ received (from clients or servers); a $proposed$ set (line \ref{line:tob.initproposed}), containing all elements of $observed$ for which $\server$ submitted its \con proposal; a $candidates$ set (line \ref{line:tob.initcandidates}), containing all elements of $observed$ that qualify as candidates (see \cref{subsection:flutter.overview}); a $delivered$ set (line \ref{line:tob.initdelivered}), tracking which $(client, message)$ tuples $\server$ already delivered (this ensures deduplication should a client, \eg attempt broadcasting the same message with several different bets); a $remote\_times$ map (line \ref{line:tob.initremotetimes}), tracking the highest local time each server ever announced reaching; a $decisions$ map (line \ref{line:tob.initdecisions}), eventually recording \con's decision for each element of $observed$; a $last\_processed$ tuple (line \ref{line:tob.initlastprocessed}), tracking the last element of $candidates$ that was delivered or rejected ($last\_processed$ sequentially takes the value of every decided candidate, see \cref{subsection:flutter.overview}).

\paragraph{Time.}
Procedure $beat$ (line \ref{line:tob.beat}) disseminates $\server$'s local time to all servers using a $\texttt{Time}$ message (lines \ref{line:tob.sendtimefor} and \ref{line:tob.sendtime}).
Upon delivering a $\texttt{Time}$ message (line \ref{line:tob.delivertime}), $\server$ updates the relevant entry in $remote\_times$ (line \ref{line:tob.updateremotetimes}, note the use of $\max$ to ensure that every value in $remote\_times$ is non-decreasing).
Procedure $lock\_time$ (line \ref{line:tob.locktime}, see \cref{subsection:flutter.overview}) returns the highest local time that at least $4 \faultcount + 1$ announced reaching, as recorded in $remote\_times$ (line \ref{line:tob.locktimereturn}).

\paragraph{Dissemination.}
Procedure $spot$ (line \ref{line:tob.spot}) handles the discovery and propagation of new client-issued messages.
Upon executing $spot\rp{\client, m, b}$: if $b$ is greater than $\server$'s lock time, $\server$ marks $\rp{\client, m, b}$ as a candidate (lines \ref{line:tob.candidatecondition} and \ref{line:tob.addcandidate}, see \cref{subsection:flutter.overview}); if $\server$ never observed $\rp{\client, m, b}$ before (\ie $\server$ never executed $spot\rp{\client, m, b}$ before, lines \ref{line:tob.spotcondition} and \ref{line:tob.addobserved}), $\server$ relays $\rp{\client, m, b}$ to every server by means of an $\texttt{Observe}$ message (lines \ref{line:tob.sendobservefor} and \ref{line:tob.sendobserve}), then schedules $beat()$ for execution at time $b$ (line \ref{line:tob.schedulebeat}) --- $\server$ does so to ensure that every correct server's lock time eventually reaches $b$, enabling the delivery of $m$ should \con decide that $m$ should be delivered (see \cref{subsection:flutter.overview}).
Upon receiving an $\texttt{Observe}$ message (line \ref{line:tob.deliverobserve}) $\server$ invokes $spot$ (line \ref{line:tob.spotuponobserve}), thus ensuring that the message is propagated.

\begin{algorithm}[t!]
\begin{algorithmic}[1]
\Implements
    \Instance{\tobab}{\tobin}
\EndImplements
\algoblockspacing

\Uses
    \Instance{AuthenticatedFifoLinks}{af}
    \Instances{\conab}{\conin}
\EndUses
\algoblockspacing

\Upon{\tobin}{Init}{}
    \State $observed: \set{\clients \times \messages \times \altime} = \cp{}$ \label{line:tob.initobserved}
    \State $proposed: \set{\clients \times \messages \times \altime} = \cp{}$ \label{line:tob.initproposed}
    \State $candidates: \set{\clients \times \messages \times \altime} = \cp{}$ \label{line:tob.initcandidates}
    \State $delivered: \set{\clients \times \messages} = \cp{}$ \label{line:tob.initdelivered}
    \State $remote\_times: \map{\servers}{\altime} = \cp{\server \rightarrow -\infty}_{\server \in \servers}$ \label{line:tob.initremotetimes}
    \State $decisions: \map{\rp{\clients \times \messages \times \altime}}{\values} = \cp{}$ \label{line:tob.initdecisions}
    \State $last\_processed: \clients \times \messages \times \altime = \rp{\bot, \bot, -\infty}$ \label{line:tob.initlastprocessed}
\EndUpon
\algoblockspacing

\Procedure{beat}{{}} \label{line:tob.beat}
    \ForAll{\server}{\servers} \label{line:tob.sendtimefor}
        \Trigger{af}{Send}{\qp{\texttt{Time}, local\_time\rp{}}} \label{line:tob.sendtime}
    \EndForAll
\EndProcedure
\algoblockspacing

\Upon{af}{Deliver}{\server, \qp{\texttt{Time}, t}} \label{line:tob.delivertime}
    \State $remote\_times\qp{\server} = \max\rp{remote\_times\qp{\server}, t}$ \label{line:tob.updateremotetimes}
\EndUpon
\algoblockspacing

\Procedure{lock\_time}{{}} \label{line:tob.locktime}
    \State $\return \max t \such \abs{\cp{\server \such remote\_times\qp{\server} \geq t}} \geq 4 \faultcount + 1$ \label{line:tob.locktimereturn}
\EndProcedure
\algoblockspacing

\Procedure{spot}{\client, m, b} \label{line:tob.spot}
    \If{$b > lock\_time\rp{}$} \label{line:tob.candidatecondition}
        \State $candidates \leftarrow \rp{\client, m, b}$ \label{line:tob.addcandidate}
    \EndIf
    \If{$(\client, m, b) \notin observed$} \label{line:tob.spotcondition}
        \ForAll{\server}{\servers} \label{line:tob.sendobservefor}
            \Trigger{af}{Send}{\qp{\texttt{Observe}, \client, m, b}} \label{line:tob.sendobserve}
        \EndForAll
        \Schedule{beat()}{b} \label{line:tob.schedulebeat}
        \State $observed \leftarrow \rp{\client, m, b}$ \label{line:tob.addobserved}
    \EndIf
\EndProcedure
\algoblockspacing

\Upon{af}{Deliver}{\server, \qp{\texttt{Observe}, \client, m, b}} \label{line:tob.deliverobserve}
    \State $spot\rp{\client, m, b}$ \label{line:tob.spotuponobserve}
\EndUpon
\algoblockspacing

\Upon{af}{Deliver}{\client, \qp{\texttt{Message}, m, b}} \label{line:tob.delivermessage}
    \State $spot(\client, m, b)$ \label{line:tob.spotuponmessage}
    \If{$\rp{\client, m, b} \notin proposed$} \label{line:tob.proposedcheckuponmessage}
        \State $in\_time: \bool = \rp{b > local\_time\rp{}}$ \label{line:tob.intime}
        \Trigger{\conin\qp{\rp{\client, m, b}}}{Propose}{in\_time} \label{line:tob.proposeuponmessage}
        \State $proposed \leftarrow \rp{\client, m, b}$ \label{line:tob.addproposeduponmessage}
    \EndIf
\EndUpon
\algoblockspacing

\UponExists{\rp{\client, m, b} \in \rp{observed \setminus proposed}\newline}{b \leq local\_time\rp{}} \label{line:tob.proposeexpirecondition}
    \Trigger{\conin\qp{\rp{\client, m, b}}}{Propose}{\false} \label{line:tob.proposeuponexpire}
    \State $proposed \leftarrow \rp{\client, m, b}$ \label{line:tob.addproposeduponexpire}
\EndUponExists
\algoblockspacing

\Upon{\conin\qp{\rp{\client, m, b}}}{Decide}{v} \label{line:tob.decide}
    \Trigger{af}{Send}{\client, \qp{\texttt{Decision}, m, b, v}} \label{line:tob.senddecision}
    \State $decisions\qp{\rp{\client, m, b}} = v$ \label{line:tob.setdecision}
\EndUpon
\algoblockspacing

\UponCondition{\min \rp{\client, m, b} \in candidates\newline\algst \rp{\client, m, b} > last\_processed\newline\satisfies \rp{\client, m, b} \in decisions \algand b \leq lock\_time\rp{}} \label{line:tob.processcondition}
    \If{$decisions\qp{\rp{\client, m, b}} = \true$} \label{line:tob.ordercondition}
        \State $order\rp{\client, m, b}$ \label{line:tob.invokeorder}
    \EndIf
    \State $last\_processed = \rp{\client, m, b}$ \label{line:tob.updatelastprocessed}
\EndUponCondition
\algoblockspacing

\Procedure{order}{\client, m, b} \label{line:tob.order}
    \If{$\rp{\client, m} \notin delivered$} \label{line:tob.delivercheck}
        \Trigger{\tobin}{Deliver}{\client, m} \label{line:tob.deliver}
        \State $delivered \leftarrow \rp{\client, m}$ \label{line:tob.adddelivered}
    \EndIf
\EndProcedure
\end{algorithmic}
\caption{\tobal server}
\label{algorithm:tobalserver}
\end{algorithm}

\paragraph{Proposals.}
Upon receiving a $\texttt{Message}$ message conveying a message $m$ with bet $b$ from client $\client$ (line \ref{line:tob.delivermessage}): $\server$ invokes $spot$ to record and propagate $\rp{\client, m, b}$ (line \ref{line:tob.spotuponmessage}); if $\server$ never proposed delivering or rejecting $m$ (lines \ref{line:tob.proposedcheckuponmessage} and \ref{line:tob.addproposeduponmessage}), $\server$ proposes delivering $m$ (line \ref{line:tob.proposeuponmessage}) if and only if $b$ is in the future according to $\server$'s local time (line \ref{line:tob.intime}, see \cref{subsection:flutter.overview}).
If $\server$'s local time reaches the bet $b$ of some message $m$ that $\server$ never proposed delivering or rejecting (lines \ref{line:tob.proposeexpirecondition} and \ref{line:tob.addproposeduponexpire}), $\server$ proposes rejecting $m$ (line \ref{line:tob.proposeuponexpire}).
This condition triggers for $\rp{\client, m, b}$ only if $\server$ discovered $\rp{\client, m, b}$ from another server's $\texttt{Observe}$ message (lines \ref{line:tob.deliverobserve}, \ref{line:tob.spotuponmessage}, \ref{line:tob.spot}, \ref{line:tob.spotcondition}, then \ref{line:tob.addobserved}), but $\server$ never received $m$ directly from $\client$ (lines \ref{line:tob.delivermessage}, \ref{line:tob.proposeuponmessage}, then \ref{line:tob.addproposeduponmessage}).
In that case, $\server$ could not authenticate $\client$ having issued $m$, and thus could not propose delivering $m$, lest $m$ be spuriously attributed to $\client$ by a malicious server.
To ensure liveness, however, every correct server must submit a proposal to every Consensus instance.
As soon as $\server$'s local time reaches $b$, $\server$ would propose to reject $m$ (lines \ref{line:tob.intime} and \ref{line:tob.proposeuponmessage}) even if $\server$ received $m$ from $\client$ (line \ref{line:tob.delivermessage}): as such, $\server$ can safely propose rejecting $m$.

\paragraph{Decision and delivery.}
Upon deciding whether to deliver a message $m$ issued by client $\client$ with bet $b$ (line \ref{line:tob.decide}), $\server$ relays the decision to $\client$ via a $\texttt{Decision}$ message (line \ref{line:tob.senddecision}), then records the decision (line \ref{line:tob.setdecision}).
$\server$ processes, \ie delivers or rejects, each candidate in strict  ascending order; $\server$ is ready to process the next candidate $\rp{\client, m, b}$ as soon as a decision is recorded for $\rp{\client, m, b}$ and $\server$'s lock time reaches $b$ (line \ref{line:tob.processcondition}).
Upon processing $\rp{\client, m, b}$, $\server$ invokes procedure $order$ (line \ref{line:tob.invokeorder}) if and only if $\server$ decided that $m$ should be delivered (line \ref{line:tob.ordercondition}).
Procedure $order$ (line \ref{line:tob.order}) delivers $\rp{\client, m}$ (line \ref{line:tob.deliver}) if $\rp{\client, m}$ was not delivered before (lines \ref{line:tob.delivercheck} and \ref{line:tob.adddelivered}).

\subsection{Correctness Sketches}
\label{subsection:tob.correctnesssketches}

This section includes proof sketches of the correctness of each of \tobal's property.
Proofs to the fullest extent of formal detail are provided in \cref{section:tob.correctness}.

\paragraph{No Duplication.}
The $delivered$ set guards the deliveries, it is extended upon delivery.

\paragraph{Integrity.}
A correct server proposes to deliver a message $m$ from a client $\client$ only upon receiving $m$ directly from $\client$.
Unless $\client$ issued $m$, no correct server ever proposes delivering $m$ and, by \con's \validity, $m$ is rejected.

\paragraph{Agreement and Total Order.}
Upon receiving a message $m$ issued by client $\client$ with bet $b$, a correct server $\server$ relays $\rp{\client, m, b}$ to every other server.
As a result, every correct server eventually observes $\rp{\client, m, b}$ (\ie adds $\rp{\client, m, b}$ to $observed$).
Moreover, if $\server$ does not submit any proposal to $\conin\qp{\rp{\client, m, b}}$ by time $b$, $\server$ eventually proposes $\false$ to $\conin\qp{\rp{\client, m, b}}$.
Hence, every correct server eventually submits a proposal to $\conin\qp{\rp{\client, m, b}}$: by the Termination and Agreement of \con, every correct server eventually records the same decision from $\conin\qp{\rp{\client, m, b}}$.
Moreover, if $\server$ does not mark $\rp{\client, m, b}$ as a candidate (\ie $\server$ does not add $\rp{\client, m, b}$ to $candidates$), then $\conin\qp{\rp{\client, m, b}}$ decides $\false$.
Indeed, if $\server$ does not mark $\rp{\client, m, b}$ as a candidate then, upon observing $\rp{\client, m, b}$, $\server$'s lock time is greater or equal to $b$.
Because links are FIFO, at least $4\faultcount + 1$ servers, of which at least $3\faultcount + 1$ are correct, announced that their local time reached $b$ before relaying $\rp{\client, m, b}$ to $\server$.
A correct server $\serversecondary$ proposes $\false$ to $\conin\qp{\rp{\client, m, b}}$ unless $\serversecondary$ observes (and relays) $\rp{\client, m, b}$ before $b$.
This proves that at least $3\faultcount + 1$ correct servers propose $\false$ to $\conin\qp{\rp{\client, m, b}}$ and, by the \validity of \con, $\conin\qp{\rp{\client, m, b}}$ decides $\false$.

Every time a correct server $\servertertiary$ observes a new message, $\servertertiary$ schedules $beat\rp{}$ for execution at some time in the future.
This means that, unless a finite number of messages is broadcast throughout the entire execution of the protocol, $\servertertiary$ executes $beat\rp{}$ an infinite number of times.
Upon executing $beat\rp{}$, $\servertertiary$ disseminates its local time to every server.
As a result, for every correct server $\servertertiary$, a correct server $\server$ indefinitely increases the value of $remote\_times\qp{\server}$.
Noting that $4\faultcount + 1$ servers are correct, this means that $\server$'s lock time grows indefinitely.
As a consequence of this, for any time $t$, $\server$ marks as candidates only a finite number of messages preceding $t$.
This is obvious if a finite number of messages is broadcast throughout the entire execution of the protocol.
If infinite messages are broadcast, $\server$ accepts new candidates preceding $t$ only until $\server$'s lock time grows past $t$ --- before that happens, $\server$ can only collect a finite number of such candidates.
Because $\server$'s candidates preceding any $t$ are always finite, $\server$'s candidates can be enumerated by a unique, strictly increasing sequence.

A correct server $\server$ processes candidates in a strictly increasing fashion.
$\server$ processes the next candidate $\rp{\client, m, b}$ once $\conin\qp{\rp{\client, m, b}}$ decides, and $\server$'s lock time reaches $b$.
This means that, once $\server$ has processed $\rp{\client, m, b}$, $\server$ will not mark any $\rp{\client', m', b'} < \rp{\client, m, b}$ as candidate.
In other words, $\server$ is guaranteed to never skip a candidate: if $\server$ was ever to mark $\rp{\client', m', b'}$ as candidate, $\server$ would have done so before processing $\rp{\client, m, b}$.
Finally, because $\server$'s candidates can be enumerated, $\server$ is guaranteed to eventually process every message it will ever mark as candidate, in strictly increasing order.
Upon processing $\rp{\client, m, b}$, $\server$ invokes $order\rp{\client, m, b}$ if and only if $\conin\qp{\rp{\client, m, b}}$ decided $\true$.
Because every $\rp{\client', m', b'}$ for which \con decides $\true$ is eventually marked by $\server$ as candidate, $\server$ invokes $order$ on a strictly increasing sequence of messages containing all and only those messages for which \con decides $\true$.
Because such sequence is unique, every correct server issues the exact same sequence of calls to $order$.
Finally, $order$ simply functions as a deterministic state machine (its state being the $delivered$ variable) to deterministically deduplicate and deliver messages.
This proves that every correct server delivers the same sequence of messages.

\paragraph{Validity.}
The implementation of a \tobal client is presented in \cref{algorithm:tobalclient} in \cref{subsection:tob.client.algorithm}.
The logic of a \tobal client is very simple: a correct client $\client$ broadcasts a message $m$ by first setting a bet that, according to $\client$'s clock, is $(\tilde \messagedelay + \betmargin)$ in the future, where $\tilde \messagedelay$ is $\client$'s best estimate for $\messagedelay$ and $\betmargin$ is an arbitrarily small constant margin.
If $m$ is rejected, $\client$ retries with a bet $2\tilde \messagedelay + \betmargin$ in the future, then $4\tilde\messagedelay + \betmargin$, and so on.
As both $\messagedelay$ and $\timedrift$ are finite, $\client$ is guaranteed to eventually select a bet $b$ such that $\rp{m, b}$ reaches each correct server before its local time reaches $b$.
When this happens, every correct server proposes delivering $m$ and, by the \validity of \con, $m$ is delivered.

\subsection{Latency Sketches}
\label{subsection:tob.latencysketches}

\begin{figure}[t]
    \centering
    \subfloat[Good-case execution.]{
        \includegraphics[height=36mm]{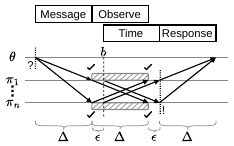}%
        \label{fig:flutter-protocol-good}%
    }
    \hfill
    \subfloat[Bad-case execution.]{
        \includegraphics[height=31mm]{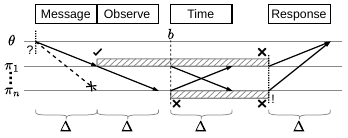}%
        \label{fig:flutter-protocol-bad}%
    }
    \caption{%
        \textbf{Examples of a good-case execution, with delivery in $2\messagedelay + \betmargin$ time units, and a bad-case execution of \tobal using \conal.}
        Notation: \tobal broadcasts are indicated with a question mark; \tobal deliveries with an exclamation mark; a correct server's executing \conal is marked by a striped bar; \conal proposals and decisions are indicated at the beginning and end of each bar with a \Checkmark for $\true$ and \textsf{X} for $\false$.
    }
    \Description[Examples of a good-case and a bad-case execution of Flutter using Blink.]{
        Both figures depict the timelines from when a message is being broadcast by a client to its delivery by all correct servers.
        In the good-case execution: all servers receive the client message in time, before its bet, hence they all propose $\true$ to the underlying instance of Blink which activates its fast path.
        Blink decides $\true$ in $\messagedelay$ time unit, leading to a rapid delivery in Flutter in $2\messagedelay + \betmargin$ time units.
        In the bad-case execution: not all servers receive the client message, hence some servers rely on other correct servers to echo the client message via an Observe.
        The minority of servers that received the client message directly propose $\true$ to Blink, the remaining majority of servers propose $\false$.
        The lack of pre-agreement activates Blink's slow path which delays the eventual decision which is $\false$, leading to a rejection of the client message.
    }
    \label{fig:flutter-protocol}%
\end{figure}

\paragraph{Good-case latency.}
We formally prove the good-case latency in \cref{subsection:tob.latency}.
\Cref{fig:flutter-protocol-good} depicts an execution of \tobal in the good case (see \cref{subsection:background.goodcase}) where clocks are synchronized and the system is synchronous, hence correct clients accurately estimate $\messagedelay$, \ie $\tilde \messagedelay = \messagedelay$.
In this example, correct client $\client$ broadcasts at time $t$ by disseminating its $\texttt{Message}$ $m$ with bet $b = t + \messagedelay + \betmargin$, with $\betmargin > 0$ an arbitrarily small constant.
By time $t + \messagedelay$, every correct server receives and supports delivering $m$ by proposing $\true$ to \con\xspace --- here, \conal.
At time $t + \messagedelay + \betmargin$, every correct server announces its local $\texttt{Time}$ being exactly $t + \messagedelay + \betmargin$.
Because every correct server has proposed the same value to \conal, every correct server fast-path decides $\true$ by time $t + 2\messagedelay$, and the $\texttt{Response}$ is sent back to $\client$.
By time $t + 2\messagedelay + \betmargin$, every correct server's lock time reaches $b = t + \messagedelay + \betmargin$, and every correct server delivers $m$.

\paragraph{Bad-case latency.}
Two classes of problem can increase the latency of a \tobal broadcast.
First, a client can increase the latency of its own broadcast by crashing or misjudging its bet.
\cref{fig:flutter-protocol-bad} depicts an example of such a scenario, where client $\client$ crashes while disseminating its $\texttt{Message}$ $m$ with bet $b \gg t + \messagedelay$.
At time $t + \messagedelay$, correct server $\server_1$ receives and relays $m$ to all other servers.
As $b > t + \messagedelay$, $\server_1$ supports delivering $m$ by proposing $\true$ to \con\xspace --- here, \conal.
At time $t + 2\messagedelay$, every correct server receives $m$ via $\server_1$'s $\texttt{Observe}$.
At time $b$, every correct server announces its local $\texttt{Time}$ and every correct server that did not receive $m$ directly from $\client$ proposes $\false$ to \conal to reject $m$.
As proposals are split, \conal's fast path cannot be used and the decision takes longer than $\messagedelay$.
Every correct server eventually decides $\false$ to reject $m$ and sends the $\texttt{Response}$ back to $\client$.

Second, a message $m$ with bet $b$ can be delayed by an independent message $m'$ with bet $b' < b$.
Consider the case where $b'$ is improperly set.
Split proposals for $m'$ push \conal out of its fast path, resulting in a delayed decision for $m'$.
If $b'$ is close enough to $b$, \conal decides to deliver $m$ while the decision for $m'$ is still pending.
Because messages are delivered in increasing order of bet, however, $m$'s delivery has to wait for $m'$'s decision.
In the worst case, this results in $m$ being delayed by one \weakcon slow path, regardless of $b$'s timeliness.
This scenario is much more severe, as even clients that are correct and timely can have their messages delayed.
As discussed in \cref{section:consensus.overview}, however, when composed with a probabilistic instance of \weakcon, \conal's worst-case latency drops in expectation to $O(1)$.
Using probabilistic \conal greatly improves the latency of a correct and timely client, as its messages can be delayed on expectation by at most a constant.

\subsection{Latency Quasi-Optimality Sketches}
\label{section:flutter.quasioptimal.sketches}

We formally prove in \cref{appendix:tob.quasioptimal} that the good-case latency of \tobal of $2 \messagedelay + \betmargin$ is quasi-optimal.
The proof assumes correct servers deliver a message in less than $2 \messagedelay$ time units, then uses undistinguishable runs to reach a contradiction.
In two correct runs with identical scheduling, a single correct client broadcasts a single message: $m$ in the first run, $m'$ in the second.
A third run is obtained as a hybrid of the first two, where the client is malicious and equivocates the message such that a server $\server$ receives $m$, while another server $\server'$ receives $m'$.
In the hybrid run, $\server$ (resp., $\server'$) observes exactly the same messages as in the first (resp., second) run.
As servers are deterministic state machines, in the hybrid run $\server$ delivers $m$ and $\server'$ delivers $m'$, breaking Agreement.

\subsection{Complexity}
\label{subsection:tob.complexity}

A broadcast attempt in \tobal involves: a client-to-servers \texttt{Message} message ($O(l)$ bits per message, where $l$ is the length of the message being broadcast); a servers-to-servers \texttt{Observe} message ($O(l)$ bits per message); a servers-to-servers \texttt{Time} message ($O(1)$ bits per message); an execution of \con (whose complexity is lower bound by $O(\servercount^2)$~\cite{dr85}); and a servers-to-client \texttt{Decision} message ($O(l)$ bits per message).
Assuming \tobal depends on an optimal implementation of \con, this results in an $O(l \servercount^2)$ complexity per broadcast attempt, which could be dominated by $l$ if the message being broadcast is large enough.
Several mitigations to this problem exist in literature, usually involving encoding the message in multiple, shorter codewords which correct servers decode after dissemination~\cite{dare-disc23,add-ccs21,dumbo-mvba-podc20,craft-fast20}.
Using one such solution, the complexity of \tobal would be reduced to $O(\servercount l + \servercount^2 \log(\servercount))$.
Optimizing \tobal's bit complexity, however, is beyond the scope of this paper.

\section{Related Work}
\label{sec:rw}

Classic Byzantine agreement protocols require more than two message delays to safely terminate.
PBFT~\cite{pbft-tocs02}, BFT-SMaRt~\cite{bft-smart-dsn14} and ICC~\cite{icc-podc22} need three rounds of one message delay each to solve Consensus in the good case.
HotStuff~\cite{hotstuff-podc19}, SQuad~\cite{squad-disc22} and Lewis-Pye's~\cite{lewispye-quadratic-arxiv22} have rounds composed of two message delays each to reduce worst-case message complexity~\cite{dr85}, increasing latency.
Leaderless approaches using DAGs showcase much higher throughput often at the cost of higher latencies: up to 12 message delays for Bullshark~\cite{bullshark-ccs22}, down to 4 for BBCA-CHAIN~\cite{bbcachain-fc24}, Autobahn~\cite{autobahn-sosp24} and Shoal++~\cite{shoalpp-nsdi25}, and 3 for Mysticeti~\cite{mysticeti-ndss25}.

These latencies do not account for the extra message delay required for a client's message to reach a server in deployments~\cite{dfinity-whitepaper-2022,aptos-whitepaper-2022,sui-whitepaper-2022}.
In comparison, \tobal's optimistic 2 message delays captures the whole picture: from client broadcast to server delivery.

\paragraph{One-message-delay Consensus.}
Multiple implementations solve crash-tolerant Consensus in one message delay in the good case~\cite{brasileiro-raynal-one-step-crash-pact01,dobre-suri-one-step-dsn06}.
The first one-message-delay Byzantine Consensus in the good case
was proposed in 2005~\cite{friedman-one-step-tdsc05}: it requires $5\faultcount + 1$ resiliency and the existence of oracles which may be implemented using a random common coin, requiring cryptographic tools, or a failure detector.
\conal's fast path does not require such an oracle.

Bosco~\cite{bosco-disc08} resembles \conal: it is Byzantine tolerant with a $7\faultcount + 1$ resiliency and enables one-message-delay latency in the good case.
Servers in Bosco operate in the stricter model of \emph{communication steps} where each server can only ``\emph{1) send messages; 2) receive messages; and 3) do local computations, in that order}''~\cite[\secref~1]{bosco-disc08}.
In comparison, servers in \conal can perform these operations in any order.
Bosco's fast path can only be used if the first $4\faultcount + 1$ values received by a server are the same, hence a single faulty server can force the slow path for all servers.
\conal's fast path activates when any set of $4\faultcount + 1$ values are identical, even if up to $\faultcount$ faulty servers propose another value.
\Cref{section:consensus.bosco} compares \conal and Bosco in detail.

\paragraph{Two-message-delay Consensus.}
Kursawe first proposed a fast Byzantine Consensus implementation~\cite{kursawe-optimistic-srds02} with two-message-delay latency and optimal $3\faultcount + 1$ resiliency in the good case, but a single failure results in all processes using the slow path, akin to Bosco.

FaB Paxos~\cite{fab-paxos-dsn05} is the first Consensus with a Byzantine fault-tolerant fast path in two message delays, with a $5\faultcount + 1$ resiliency.
Two-message-delay Byzantine Consensus was later proven solvable with an improved resiliency of $5\faultcount - 1$~\cite{optimal-fast-bft-podc21,latency-categorization-podc21}, which is optimal in small networks since $3\faultcount + 1 = 5\faultcount -1$ when $\faultcount = 1$.
Reducing the size of quorums in some of these approaches' steps decreases latency in practice without affecting the number of message delays~\cite{hu-zhang-two-rounds-ietis23}.

\paragraph{Fast \tob.}
Multiple implementations solve \emph{crash} \tob in two message delays in the good case~\cite{pedone-optimistic-tob-dc98,raynal-fast-tob-prdc00}.
SBFT~\cite{sbft-dsn19} employs a fast path in the good case for Byzantine state machine replication.
However, SBFT uses collector replicas to aggregate signatures in its rounds to increase throughput at the cost of delivery latency, akin to HotStuff.
Sync HotStuff~\cite{sync-hotstuff-sp20} reduces HotStuff's latency down to two message delays assuming synchrony and can operate with an improved $2\faultcount + 1$ resiliency.
HotStuff-2~\cite{hotstuff2-arxiv} merges two of HotStuff's three rounds resulting in a two-round approach, each taking two message delays.
VBFT~\cite{vbft-arxiv} proposes two-message-delay latency with an optimal $3\faultcount + 1$ resiliency by relaxing the safety of Consensus which may lead to temporary decision revocations.
UBFT~\cite{ubft-asplos23} uses an optimistic fast path to avoid the cost of signatures in hardware-accelerated systems.

Banyan~\cite{banyan-mw24} is the first partially-synchronous Byzantine \tob with a two-message-delay fast path when the leader broadcasts, hence three-message-delay when a client broadcasts.
Banyan defines a resiliency of $n = 3f + 2p - 1$ where $f$ is the maximum number of Byzantine faults and $p$ is freely set such that $1 \leq p \leq f$.
Banyan rapidly delivers a value if $\rp{n - p}$ processes are correct and respond fast enough.

\paragraph{Speculative execution.}
Zyzzyva and Zyzzyva5~\cite{zyzzyva-sosp07} are Byzantine replication protocols that achieve optimistic broadcast-to-delivery latencies of two message delays thanks to speculative execution.
These protocols execute client messages before agreement is reached, thus save one message delay compared to other approaches.
However, speculative execution may lead to diverging state across servers requiring a state reconciliation protocol, and its effectiveness depends on the workload, unlike in \tobal.
HotStuff-1~\cite{hotstuff1-sigmod25} reduces HotStuff-2 to a single round, of two message delays, thanks to speculative execution.
Speculative execution is an orthogonal, and possibly complementary, approach to that of \tobal.

\section{Future Work}
\label{sec:conclusion}

There are several avenues for future work on \tobal:
(1) Is there an algorithm that offers the same advantages as \tobal but only requires \weakcon instead of \con?
(2) Should such an algorithm exist, would it improve the resiliency threshold towards $3\faultcount + 1$?
(3) If not, could a trusted computing base~\cite{naama-classify-tee-podc21} be leveraged to achieve $3\faultcount + 1$ resiliency, as done in other protocols~\cite{a2m-sosp07,trinc-nsdi09,cheapbft-eurosys12,minbft-tc13,fastbft-tc19,ubft-asplos23}?
(4) How well does \tobal perform in practice considering it uses neither leaders nor signatures, and how compatible is it with hardware acceleration~\cite{consensus-box-nsdi16,fastbft-tc19,bft-smartnic-fpga-debs21,hovercraft-eurosys20}?
(5) What is the expected latency of \tobal when \conal uses a probability-1 implementation of \weakcon?

\begin{acks}
    We thank the anonymous reviewers for their constructive feedback and Yann Vonlanthen for insightful discussions on fast agreement.
This work has been funded in part by BRIDGE (\#40B1-0\_222426).

\end{acks}

\balance
\bibliographystyle{ACM-Reference-Format}
\bibliography{references.bib}

\clearpage
\appendix
\onecolumn

\tableofcontents
\label{section:toc}
\clearpage
\section{Blink}
\label{appendix:consensus}

\Cref{section:consensus.correctness} details full proofs of \conal's correctness and \cref{subsection:consensus.latency} full proofs of \conal's latency.
\Cref{section:consensus.bosco} analyzes the differences between \conal and the close algorithm Bosco~\cite{bosco-disc08}.

\subsection{Correctness}
\label{section:consensus.correctness}

\cref{subsection:consensus.termination} proves Termination.
\cref{subsection:consensus.validity} proves \validity.
\cref{subsection:consensus.integrityagreement} proves Integrity and Agreement.

\subsubsection{Termination}
\label{subsection:consensus.termination}

\begin{lemma}
\label{lemma:consensus.everyonedepdecides}
Every correct server eventually triggers $\event{\condep}{Decide}{}$.
\begin{proof}
Upon triggering $\event{\conin}{Propose}{}$ (line \ref{line:conal.propose}), every correct server sends a $\texttt{Suggest}$ message to every server (lines \ref{line:conal.suggestfor} and \ref{line:conal.sendsuggest}). 
Let $\server$ be a correct server.
Upon delivering a $\texttt{Suggest}$ message from a server $\servertertiary$ (line \ref{line:conal.deliversuggest}) $\server$ adds $\servertertiary$ to the keys of $suggestions$ (line \ref{line:conal.recordsuggestion}).
As we assume $4\faultcount + 1$ correct servers, $\server$ eventually adds at least $4\faultcount + 1$ keys to $suggestions$.
Noting that $\server$ never removes keys from $suggestions$, $\server$ eventually satisfies $\abs{suggestions} \geq 4\faultcount + 1$.

Upon initialization, $\server$ sets $proposed = \false$ (line \ref{line:conal.initproposed}); $\server$ updates $proposed$ to $\true$ (line \ref{line:conal.setproposed} only) only after triggering $\event{\condep}{Propose}{}$ (line \ref{line:conal.deppropose}).
This proves that $\server$ eventually triggers $\event{\condep}{Decide}{}$.
Indeed, let us assume for the sake of contradiction that $\server$ never triggers $\event{\condep}{Decide}{}$.
$\server$ would eventually permanently satisfy $\abs{suggestions} \geq 4 \faultcount + 1$ and $proposed = \false$.
As a result, the event at line \ref{line:conal.depproposecondition} would eventually trigger, causing $\server$ to trigger $\event{\condep}{Propose}{}$.

In summary, every correct server eventually triggers $\event{\condep}{Propose}{}$.
By the Termination property of $\condep$, every correct server eventually triggers $\event{\condep}{Decide}{}$, and the lemma is proved.
\end{proof}
\end{lemma}

\begin{theorem}
\label{theorem:consensus.termination}
\conal satisfies Termination.

\begin{proof}
Upon initialization, a correct server sets $decided = \false$ (line \ref{line:conal.initdecided}). 
A correct server updates $decided$ to $\true$ (lines \ref{line:conal.setdecidedfast} and \ref{line:conal.setdecidedslow} only) only after triggering $\event{\conin}{Decide}{}$ (lines \ref{line:conal.decidefast} and \ref{line:conal.decideslow}).
Moreover, by Lemma \ref{lemma:consensus.everyonedepdecides}, every correct server eventually triggers $\event{\condep}{Decide}{}$.
Upon doing so (line \ref{line:conal.depdecide}), a correct server either satisfies $decided = \true$, in which case it previously triggered $\event{\conin}{Decide}{}$, or $decided = \false$ (line \ref{line:conal.checkdecidedslow}), in which case it immediately triggers $\event{\conin}{Decide}{}$ (line \ref{line:conal.decideslow}).
In summary, every correct server eventually triggers $\event{\conin}{Decide}{}$, and the theorem is proved.
\end{proof}
\end{theorem}

\subsubsection{Representative Validity}
\label{subsection:consensus.validity}

\begin{observation}
\label{observation:consensus.suggestmessages}
Let $\server$ be a correct server.
Upon initialization, $\server$ sets $suggestions = \cp{}$ (line \ref{line:conal.initsuggestions}).
$\server$ sets $suggestions\qp{\servertertiary} = v$ (line \ref{line:conal.recordsuggestion} only) only upon delivering a $\qp{\texttt{Suggest}, v}$ message from $\servertertiary$ (line \ref{line:conal.deliversuggest}).
This means that, if $suggestions\qp{\servertertiary} = v$ at $\server$, then $\server$ delivered a $\qp{\texttt{Suggest}, v}$ message from $\servertertiary$.
\end{observation}

\begin{lemma}
\label{lemma:consensus.fastconditionimpliesmajority}
Let $v \in \values$. If any correct server eventually satisfies
\begin{equation}
\label{equation:consensus.fastconditionimpliesmajority.fastcondition}
    \abs{\cp{\servertertiary \in \servers \such suggestions\qp{\servertertiary} = v}} \geq 4\faultcount + 1
\end{equation}
then no correct server ever satisfies
\begin{equation}
\label{equation:consensus.fastconditionimpliesmajority.majority}
    \abs{\cp{\servertertiary \in \servers \such suggestions\qp{\servertertiary} = \bar v}} \geq 2\faultcount + 1
\end{equation}
\begin{proof}
Let us assume, for the sake of contradiction, that two correct servers $\server, \serversecondary$ exist such that $\server$ satisfies Equation \ref{equation:consensus.fastconditionimpliesmajority.fastcondition} and $\serversecondary$ satisfies Equation \ref{equation:consensus.fastconditionimpliesmajority.majority}.
Let $\serverssubset_\server$, $\serverssubset_\serversecondary$ respectively denote $\server$'s suggestions for $v$ and $\serversecondary$'s suggestions for $\bar v$:
\begin{align*}
    \serverssubset_\server &= \cp{\servertertiary \in \servers \such suggestions\qp{\servertertiary} = v \text{ at } \server} \\
    \serverssubset_\serversecondary &= \cp{\servertertiary \in \servers \such suggestions\qp{\servertertiary} = \bar v \text{ at } \serversecondary}
\end{align*}
Note that $\abs{\serverssubset_\server} = 4 \faultcount + 1$, $\abs{\serverssubset_\serversecondary} = 2 \faultcount + 1$.
As we assume at most $f$ faulty servers, $\serverssubset_\server$ and $\serverssubset_\serversecondary$ respectively contain $3\faultcount + 1$ and $\faultcount + 1$ servers that are certainly correct.
This proves that at least one correct server $\servertertiary$ belongs to both $\serverssubset_\server$ and $\serverssubset_\serversecondary$.

By Observation \ref{observation:consensus.suggestmessages}, for $\servertertiary$ to belong to both $\serverssubset_\server$ and $\serverssubset_\serversecondary$, $\servertertiary$ must have sent a $\qp{\texttt{Suggest}, v}$ message to $\server$, as well as a $\qp{\texttt{Suggest}, \bar v}$ message to $\serversecondary$.
$\servertertiary$ issues $\texttt{Suggest}$ messages only once (lines \ref{line:conal.suggestfor} and \ref{line:conal.sendsuggest}), upon triggering $\event{\conin}{Propose}{x}$ (line \ref{line:conal.propose}).
Upon doing so, $\servertertiary$ sends the same message $\qp{\texttt{Suggest}, x}$ to every server.
This contradicts $\servertertiary$ sending $\qp{\texttt{Suggest}, v}$ to $\server$ and $\qp{\texttt{Suggest}, \bar v}$ to $\serversecondary$ and proves the lemma.
\end{proof}
\end{lemma}

\begin{lemma}
\label{lemma:consensus.fastconditionimpliesdecision}
Let $v \in \values$. If any correct server eventually satisfies
\begin{equation}
    \abs{\cp{\servertertiary \in \servers \such suggestions\qp{\servertertiary} = v}} \geq 4\faultcount + 1
\end{equation}
then every correct server that triggers $\event{\condep}{Decide}{}$ triggers $\event{\condep}{Decide}{v}$.
\begin{proof}
By Lemma \ref{lemma:consensus.fastconditionimpliesmajority}, no correct server satisfies
\begin{equation*}
    \abs{\cp{\servertertiary \in \servers \such suggestions\qp{\servertertiary} = \bar v}} \geq 2\faultcount + 1
\end{equation*}
This means that no correct server ever triggers $\event{\condep}{Propose}{\bar v}$.
Indeed, a correct server triggers $\event{\condep}{Propose}{x}$ (line \ref{line:conal.deppropose} only) only if $x$ satisfies $\abs{\cp{\servertertiary \in \servers \such suggestions\qp{\servertertiary} = x}} \geq 2\faultcount + 1$ (line \ref{line:conal.suggestionmajority}).
By the \weakvalidity property of $\condep$, no correct server triggers $\event{\condep}{Decide}{\bar v}$, which proves the lemma.
\end{proof}
\end{lemma}

\begin{lemma}
\label{lemma:consensus.decisionconsistency}
If a correct server triggers both $\event{\condep}{Decide}{v}$ and $\event{\conin}{Decide}{w}$, then $v = w$.
\begin{proof}
Let $\server$ be a correct server that triggers both $\event{\condep}{Decide}{v}$ (line \ref{line:conal.depdecide}) and $\event{\conin}{Decide}{w}$ (line \ref{line:conal.decidefast} or \ref{line:conal.decideslow}).
Upon triggering $\event{\condep}{Decide}{v}$ (line \ref{line:conal.depdecide}), $\server$ satisfies either $decided = \false$ or $decided = \true$.
If $decided = \false$ (line \ref{line:conal.checkdecidedslow}), then $\server$ immediately triggers $\event{\conin}{Decide}{v}$ (line \ref{line:conal.decideslow}), hence $w = v$.
Throughout the remainder of this proof, we assume that, upon triggering $\event{\condep}{Decide}{v}$, $\server$ satisfies $decided = \true$.

Upon initialization, $\server$ sets $decided = \false$ (line \ref{line:conal.initdecided}).
By the Integrity property of $\condep$, $\server$ does not trigger $\event{\condep}{Decide}{}$ more than once.
As a result, $\server$ cannot have previously set $decided = \false$ by executing line \ref{line:conal.setdecidedslow} - the line is guarded by $\event{\condep}{Decide}{}$ (line \ref{line:conal.depdecide}).
The only other option is that $\server$ previously set $decided = \false$ by executing line \ref{line:conal.setdecidedfast}.
To do so, $\server$ must have satisfied
\begin{equation*}
    \abs{\cp{\servertertiary \in \servers \such suggestions\qp{\servertertiary} = w}} \geq 4\faultcount + 1
\end{equation*}
(line \ref{line:conal.fastcondition}), as $\server$ also immediately triggers $\event{\conin}{Decide}{w}$ (line \ref{line:conal.decidefast}).
By Lemma \ref{lemma:consensus.fastconditionimpliesdecision}, however, $\server$ can only trigger $\event{\condep}{Decide}{w}$, which proves $v = w$ and concludes the lemma.
\end{proof}
\end{lemma}

\begin{theorem}
\label{theorem:consensus.validity}
\conal satisfies \validity.
\begin{proof}
Let us assume that some correct server $\server$ triggers $\event{\conin}{Decide}{v}$ for some $v \in \values$.
By Lemma \ref{lemma:consensus.everyonedepdecides}, $\server$ eventually triggers $\event{\condep}{Decide}{}$.
By Lemma \ref{lemma:consensus.decisionconsistency}, $\server$ triggers $\event{\condep}{Decide}{v}$.
By the \weakvalidity property of $\condep$, at least one correct server $\serversecondary$ triggered $\event{\condep}{Propose}{v}$.
$\serversecondary$ does so (line \ref{line:conal.deppropose} only) only if $\serversecondary$ satisfies
\begin{equation*}
    \abs{\cp{\servertertiary \in \servers \such suggestions\qp{\servertertiary} = \bar v}} \geq 2\faultcount + 1
\end{equation*}
By Observation \ref{observation:consensus.suggestmessages}, $\serversecondary$ delivered a $\qp{\texttt{Suggest}, v}$ message from $2\faultcount + 1$ distinct servers.
As we assume at most $\faultcount$ faulty servers, at least $\faultcount + 1$ correct servers issued a $\qp{\texttt{Suggest}, v}$ message.

A correct server issues a $\qp{\texttt{Suggest}, v}$ message (line \ref{line:conal.sendsuggest} only) only upon triggering $\event{\conin}{Propose}{v}$ (line \ref{line:conal.propose}).
This proves that at least $\faultcount + 1$ correct servers triggered $\event{\conin}{Propose}{v}$ and concludes the theorem.
\end{proof}
\end{theorem}

\subsubsection{Integrity and Agreement}
\label{subsection:consensus.integrityagreement}

\begin{theorem}
\label{theorem:consensus.integrity}
\conal satisfies Integrity.

\begin{proof}
A correct server triggers $\event{\conin}{Decide}{}$ (line \ref{line:conal.decidefast} or \ref{line:conal.decideslow}) only if $decided = \false$ (lines \ref{line:conal.checkdecidedfast} and \ref{line:conal.checkdecidedslow}).
Immediately after triggering $\event{\conin}{Decide}{}$, a correct server sets $decided = \true$ (line \ref{line:conal.setdecidedfast} or \ref{line:conal.setdecidedslow}).
A correct server never sets $decided$ back to $\false$.
Hence, a correct server triggers $\event{\conin}{Decide}{}$ at most once, and the theorem is proved.
\end{proof}
\end{theorem}

\begin{theorem}
\label{theorem:consensus.agreement}
\conal satisfies Agreement.
\begin{proof}
By Lemma \ref{lemma:consensus.everyonedepdecides}, every correct server eventually triggers $\event{\condep}{Decide}{}$.
By the Agreement property of $\condep$, every correct server eventually triggers $\event{\condep}{Decide}{v}$ for the same value $v$.
By Lemma \ref{lemma:consensus.decisionconsistency}, every correct server that triggers $\event{\conin}{Decide}{}$ triggers $\event{\conin}{Decide}{v}$, and the theorem is proved
\end{proof}
\end{theorem}

\subsection{Latency}
\label{subsection:consensus.latency}

\begin{theorem}
\label{theorem:consensus.latency}
If every correct server triggers $\event{\conin}{Propose}{v}$ for the same $v \in \values$, every correct server triggers $\event{\conin}{Decide}{}$ within one message delay.
\begin{proof}
Let $v \in \values$, let us assume that every correct server triggers $\event{\conin}{Propose}{v}$.
Let $\serversecondary$ be a correct server.
Upon triggering $\event{\conin}{Propose}{v}$ (line \ref{line:conal.propose}) $\serversecondary$ sends a $\qp{\texttt{Suggest}, v}$ message to every server (lines \ref{line:conal.suggestfor} and \ref{line:conal.sendsuggest}).
Because $\serversecondary$ never issues a $\texttt{Suggest}$ message other than upon triggering $\event{\conin}{Propose}{}$, and because $\serversecondary$ never triggers $\event{\conin}{Propose}{}$ more than once, $\serversecondary$ never issues any $\qp{\texttt{Suggest}, \bar v}$ message.

Let $\server$ be a correct server.
As we assume at most $\faultcount$ faulty servers, within one message delay $\server$ delivers at least $4 \faultcount + 1$ $\qp{\texttt{Suggest}, v}$ messages from correct sources.
Upon delivering a $\qp{\texttt{Suggest}, v}$ message from a correct server $\servertertiary$ (line \ref{line:conal.deliversuggest}), $\server$ sets $suggestions\qp{\servertertiary} = v$ (line \ref{line:conal.recordsuggestion}).
Because $\servertertiary$ never issues a $\qp{\texttt{Suggest}, \bar v}$ message, and because $\server$ sets $suggestions\qp{\servertertiary} = x$ (line \ref{line:conal.recordsuggestion} only) only upon delivering a $\qp{\texttt{Suggest}, x}$ message from $\servertertiary$ (line \ref{line:conal.deliversuggest}), within one message delay $\server$ satisfies
\begin{equation*}
    \abs{\cp{\servertertiary \in \servers \such suggestions\qp{\servertertiary} = v}} \geq 4 \faultcount + 1
\end{equation*}
and triggers the event at line \ref{line:conal.fastcondition}.

Upon initialization, $\server$ sets $decided = \false$ (line \ref{line:conal.initdecided}).
$\server$ updates $decided$ to $\true$ (lines \ref{line:conal.setdecidedfast} and \ref{line:conal.setdecidedslow} only) only after triggering $\event{\conin}{Decide}{}$ (lines \ref{line:conal.decidefast} and \ref{line:conal.decideslow}).
Upon triggering the event at line \ref{line:conal.fastcondition}, $\server$ satisfies either $decided = \true$ or $decided = \false$.
If $decided = \true$, then $\server$ previously triggered $\event{\conin}{Decide}{}$.
If $decided = \false$ (line \ref{line:conal.checkdecidedfast}) then $\server$ triggers $\event{\conin}{Decide}{}$ (line \ref{line:conal.decidefast}).
This proves that $\server$ triggers $\event{\conin}{Decide}{}$ within one message delay and concludes the theorem.
\end{proof}
\end{theorem}

\subsection{Comparison with Bosco}
\label{section:consensus.bosco}

The goal of this section is twofold.
First, we compare \conal with Bosco, the algorithm presented by Song and van Renesse at DISC '08~\cite{bosco-disc08}, which \conal generalizes to the $5 \faultcount + 1$ setting.
Second, we address the impossibility result, presented in the same paper, stating that $5\faultcount + 1$ Consensus cannot be solved in one communication step if any server is faulty.

\paragraph{\conal vs. Bosco.}
Like \conal, Bosco~\cite[Algorithm 1]{bosco-disc08} has every correct server disseminate its proposal (line 1) by means of a \texttt{Vote} message (akin to \conal's \texttt{Suggest} message).
Also like \conal, Bosco has every correct server collect $\servercount - \faultcount$ \texttt{Vote} messages (line 2), then propose to an underlying instance of Consensus (line 8) the \emph{majority value}, \ie the value (if it exists) that received more than $\rp{\servercount - \faultcount} / 2$ votes (lines 5 to 8).
A majority value is guaranteed to exist for \conal (which solves \weakcon), but not for Bosco (which solves multi-valued Consensus) --- in Bosco, if a correct server does not observe any value having the majority of votes, the server proposes its original proposal.
Unlike \conal, however, Bosco's fast-path decision (lines 3 and 4) can only occur at the same time as the proposal to the underlying Consensus, \ie as soon as $\servercount - \faultcount$ \texttt{Vote} messages are collected (line 2).
By contrast, a correct \conal server keeps collecting \texttt{Suggest} messages even after proposing to Consensus: if, at \emph{any} time, either value collects $4\faultcount + 1$ suggestions, the server decides that value.
Due to this difference, Bosco cannot guarantee a one-message-delay decision in the $5 \faultcount + 1$ setting, even if all correct servers propose the same value.
In the $5 \faultcount + 1$ setting, a correct Bosco server would fast-path decide only if all the first $4\faultcount + 1$ votes it received were for the same value.
Even assuming all correct servers propose the same value, the arbitrary vote of a single faulty server would be sufficient to inhibit Bosco's fast path.
That is not the case in \conal: regardless of the arbitrary behavior of faulty processes, if every correct server proposes the same value, every correct server collects $4 \faultcount + 1$ matching suggestions and fast-path decides within one message delay.

\paragraph{Message delays vs. communication steps.}
Importantly, the goal of Bosco is not to enable fast-path termination within one \emph{message delay}, but rather within one \emph{communication step}.
The difference is subtle but noteworthy.
Bosco abides by a model of asynchronous networking~\cite{latency-consensus-report01} that defines a communication step as a period of time where each process can (1) send messages; (2) receive messages; and (3) do local computations, in that order.
Conversely, this paper denotes a message delay as the bound $\messagedelay$ on communication delay (see \cref{subsection:background.model}).
In the $7 \faultcount + 1$ setting, and assuming all correct servers propose the same value, Bosco does ensure one-step termination (with its thresholds adapted to the $7 \faultcount + 1$ setting, \conal would do the same).
Remarkably, the paper proves~\cite[Theorem 1]{bosco-disc08} that Bosco is optimal in that regard, meaning one-step termination cannot be ensured by any algorithm with less than $7\faultcount + 1$ servers.
\conal's one-message-delay termination does not conflict with this result.
Indeed, \conal does not, in general, terminate within one communication step.
Consider the case where all correct servers propose the same value $v$.
If even a single faulty server issues a $\qp{\texttt{Suggest}, \bar v}$ message, the first $4 \faultcount + 1$ suggestions a correct server $\server$ collects are by no means guaranteed to be all for $v$.
Upon triggering the event at line \ref{line:conal.depproposecondition}, $\server$ would thus engage in local computation, exhausting its first step of communication without deciding $v$.

\clearpage
\section{Flutter}
\label{appendix:tob}

\Cref{subsection:tob.client.algorithm} first presents \tobal's client algorithm.
\Cref{subsection:tob.auxiliary} presents auxiliary results and notations required for the following sections.
\Cref{section:tob.correctness} contains proofs to the fullest extent of formal details of \tobal's correctness.
\Cref{subsection:tob.latency} fully proves the good-case latency of \tobal, and \cref{appendix:tob.quasioptimal} proves that it is quasi-optimal.
\Cref{subsection:tob.dos} discusses the impact of denial of services attacks on \tobal.

\subsection{Client Algorithm}
\label{subsection:tob.client.algorithm}

\Cref{algorithm:tobalclient} describes \tobal's client algorithm.
It complements \cref{algorithm:tobalserver} that presents \tobal's server algorithm discussed in \cref{subsection:flutter.algorithm}.

\begin{algorithm}[h]
\begin{algorithmic}[1]
\Implements
    \Instance{\tobab}{\tobin}
\EndImplements
\algoblockspacing

\Uses
    \Instance{AuthenticatedFifoLinks}{af}
\EndUses
\algoblockspacing

\Parameters
    \State $\tilde \messagedelay \in \altime$: message delay estimate \Comment{Equals $\messagedelay$ under synchrony, arbitrary otherwise}
    \State $\betmargin \in \altime^+$: bet margin \Comment{Arbitrarily small}\label{line:tob.betmargin}
\EndParameters
\algoblockspacing

\Upon{\tobin}{Init}{}
    \State $submissions: \map{\messages}{\rp{\mathbb{N} \times \altime}} = \cp{}$ \label{line:tobc.initsubmissions}
    \State $decisions: \map{\rp{\messages \times \altime \times \servers}}{\values}$
\EndUpon
\algoblockspacing

\Procedure{submit}{m, r} \label{line:tobc.submit}
    \State $b = local\_time\rp{} + 2^r \tilde \messagedelay + \betmargin$ \label{line:tobc.bet} \Comment{Exponential backoff}
    \ForAll{\server}{\servers} \label{line:tobc.sendmessagefor}
        \Trigger{af}{Send}{\server, \qp{\texttt{Message}, m, b}} \label{line:tobc.sendmessage}
    \EndForAll
    \State $submissions\qp{m} = (r, b)$ \label{line:tobc.setsubmission}
\EndProcedure
\algoblockspacing

\Upon{\tobin}{Broadcast}{m} \label{line:tobc.broadcast}
    \State $submit\rp{m, 0}$ \label{line:tobc.submituponbroadcast}
\EndUpon
\algoblockspacing

\Upon{af}{Deliver}{\server, \qp{\texttt{Decision}, m, b, v}} \label{line:tobc.deliverdecision}
    \State $decisions\qp{\rp{m, b, \server}} = v$ \label{line:tobc.setdecision}
\EndUpon
\algoblockspacing

\UponExists{\rp{m \rightarrow \rp{r, b}} \in submissions}{\abs{\cp{\server \such decisions\qp{\rp{m, b, \server}} = \false}} \geq \faultcount + 1} \label{line:tobc.failurecondition}
    \State $submit\rp{m, r + 1}$ \label{line:tobc.submituponfailure}
\EndUponExists
\end{algorithmic}
\caption{\tobal client}
\label{algorithm:tobalclient}
\end{algorithm}

\subsection{Notation and Auxiliary Results}
\label{subsection:tob.auxiliary}

\begin{notation}[Set crop]
Let $X$ be a totally ordered set, let $x \in X$. We use
\begin{align*}
    \crop{X}{\leq x} &= \cp{y \in X \such y \leq x} \\
    \crop{X}{<x} &= \cp{y \in X \such y < x}
\end{align*}
\end{notation}

\begin{notation}[Range]
Let $a, b \in \mathbb{N}$ such that $a \leq b$. We use
\begin{equation*}
    a..b = \cp{a, a + 1, \ldots, b}
\end{equation*}
\end{notation}

\begin{lemma}
\label{lemma:tob.enumerateset}
Let $X$ be a totally ordered set such that
\begin{equation*}
    \forall x \in X, \; \abs{\crop{X}{\leq x}} < \infty
\end{equation*}
A strictly increasing sequence $z$ exists such that
\begin{equation*}
    \forall x \in X, \; \exists n \such z_n = x
\end{equation*}
\begin{proof}
Let $\phi: X \rightarrow \mathbb{N}$ be defined by
\begin{equation*}
    \phi\rp{x} = \abs{\crop{X}{\leq x}}
\end{equation*}
The lemma is proved by showing that $\phi$ is a bijection between $X$ and $\cp{n \in \mathbb{N} \such n \leq \abs{X}}$, and that $\phi^{-1}$ is strictly increasing.

\begin{lresult}
\label{result:tob.enumerateset.inclusions}
Let $x, y \in X$ such that $x < y$. We have $\crop{X}{\leq x} \subset \crop{X}{\leq y}$.
\begin{proof}
Let $h \in \crop{X}{\leq x}$. 
We have $h \leq x < y$, which proves $h \in \crop{X}{\leq y}$.
Moreover, because $x < y$, we have $y \notin \crop{X}{\leq x}$.
Finally, because $y \leq y$, we obviously have $y \in \crop{X}{\leq y}$.
In summary, $\crop{X}{\leq x} \subseteq \crop{X}{\leq y}$ and $\crop{X}{\leq y} \setminus \crop{X}{\leq x} \supseteq \cp{y} \neq \emptyset$, hence $\crop{X}{\leq x} \subset \crop{X}{\leq y}$.
\end{proof}
\end{lresult}

\begin{lresult}
\label{result:tob.enumerateset.injective}
$\phi$ is injective.
\begin{proof}
Let $x, y \in X$ such that $x \neq y$.
Let us assume without loss of generality $x < y$.
By Result \ref{result:tob.enumerateset.inclusions}, we have
\begin{equation*}
    \phi\rp{x} = \abs{\crop{X}{\leq x}} < \abs{\crop{X}{\leq y}} = \phi\rp{y}
\end{equation*}
which proves $\phi\rp{x} \neq \phi\rp{y}$.
\end{proof}
\end{lresult}

\begin{lresult}
\label{result:tob.enumerateset.pigeonhole}
Let $x \in X$. For all $n \in 1..\phi(x)$, some $h \in X$ exists such that $\phi\rp{h} = n$.
\begin{proof}
We start by noting that the result obviously holds true for $n = \phi\rp{x}$.

Let $y \in \crop{X}{<x}$.
Noting that $y \leq y$, we obviously have $\crop{X}{\leq y} \supseteq \cp{y}$, hence $\phi\rp{y} \geq 1$.
Moreover, because $y < x$, by Result \ref{result:tob.enumerateset.inclusions} we have $\phi\rp{y} < \phi\rp{x}$.
In summary, for every $y \in \crop{X}{<x}$, we have $\phi\rp{y} \in 1..\rp{\phi\rp{x} - 1}$.

Noting that
\begin{equation*}
    \crop{X}{<x} = \cp{y \in X \such y < x} = \cp{y \in X \such y \leq x, y \neq x} = \cp{y \in X \such y \leq x} \setminus \cp{x} = \crop{X}{\leq x} \setminus \cp{x}
\end{equation*}
we have $\abs{\crop{X}{<x}} = \phi\rp{x} - 1$.

In summary, $\abs{\crop{X}{<x}}$ contains $\phi\rp{x} - 1$ elements.
For every $y \in \crop{X}{<x}$ we have $\phi\rp{y} \in 1..\rp{\phi\rp{x} - 1}$.
For every $y, w \in \crop{X}{<x}$ such that $y \neq w$, we have $\phi\rp{y} \neq \phi\rp{w}$.
By the pigeonhole principle, for all $n \in 1..\rp{\phi\rp{x} - 1}$ some $h \in \crop{X}{<x}$ exists such that $\phi\rp{h} = n$, and the result is proved.
\end{proof}
\end{lresult}

\begin{lresult}
\label{result:tob.enumerateset.finitesurjective}
If $X$ is finite, then $\phi$ is surjective on $1..\abs{X}$.
\begin{proof}
Because $X$ is finite, $X$ has a maximum.
Let $x = \max X$.
By definition we have $\phi\rp{x} = \abs{X}$, and the result follows immediately from Result \ref{result:tob.enumerateset.pigeonhole}.
\end{proof}
\end{lresult}

\begin{lresult}
\label{result:tob.enumerateset.infinitesurjective}
If $X$ is infinite, then $\phi$ is surjective on $\mathbb{N}$.
\begin{proof}
By hypothesis, for all $x \in X$ we have $\abs{\crop{X}{\leq x}} < \infty$.
As a result we have
\begin{equation*}
    \forall x \in X, \; \rp{X \setminus \crop{X}{\leq x}} \neq \emptyset
\end{equation*}
This proves the existence of some function $\alpha: X \rightarrow X$ such that
\begin{equation*}
    \alpha\rp{x} \in \rp{X \setminus \crop{X}{\leq x}}
\end{equation*}
By definition, for all $x \in X$ we have $\alpha(x) > x$.
By Result \ref{result:tob.enumerateset.inclusions}, we then have
\begin{equation*}
    \phi\rp{\alpha\rp{x}} > \phi\rp{x}
\end{equation*}
or equivalently
\begin{equation*}
    \phi\rp{\alpha\rp{x}} \geq \phi\rp{x} + 1
\end{equation*}
Noting that, for all $x \in X$, we have $x \leq x$ hence $\phi\rp{x} \geq 1$, by induction we have
\begin{equation*}
    \forall x \in X, \forall n \in \mathbb{N}, \; \phi\rp{\alpha^n\rp{x}} \geq n
\end{equation*}

We can now prove the surjectivity of $\phi$.
Let $n \in \mathbb{N}$.
Let $x \in X$, let $y = \alpha^n\rp{x}$, we have $\phi\rp{y} \geq n$.
By Result \ref{result:tob.enumerateset.pigeonhole}, some $h \in X$ exists such that $\phi\rp{h} = n$.
Because this holds for every $n$, $\phi$ is surjective on $\mathbb{N}$.
\end{proof}
\end{lresult}

By Results \ref{result:tob.enumerateset.injective}, \ref{result:tob.enumerateset.finitesurjective} and \ref{result:tob.enumerateset.infinitesurjective}, $\phi$ is a bijection between $X$ and $\cp{n \in \mathbb{N} \such n \leq \abs{X}}$.

\begin{lresult}
\label{result:tob.enumerateset.increasing}
Let $n, m \leq \abs{X}$ such that $n < m$.
We have $\phi^{-1}\rp{n} < \phi^{-1}\rp{m}$.
\begin{proof}
For the sake of readability, let $x = \phi^{-1}\rp{n}$, let $y = \phi^{-1}\rp{m}$.
Note that, by the injectivity of $\phi^{-1}$, we have $x \neq y$.
Let us assume for the sake of contradiction that $x > y$.
By Result \ref{result:tob.enumerateset.inclusions} we would have
\begin{equation*}
    n = \phi\rp{x} > \phi\rp{y} = m
\end{equation*}
which contradicts $n < m$ and proves the result.
\end{proof}
\end{lresult}

By Result \ref{result:tob.enumerateset.increasing}, $\phi^-1$ is increasing, and the lemma is proved.

\end{proof}
\end{lemma}

\begin{lemma}
\label{lemma:tob.enumerationsareunique}
Let $X$ be a totally ordered set.
Let $z, z'$ be strictly increasing sequences such that
\begin{align*}
    \forall x \in X, \; &\exists n \such z_n = x \\
    \forall x \in X, \; &\exists n' \such z'_{n'} = x
\end{align*}
We have $z = z'$.
\begin{proof}
Let us assume for the sake of contradiction that $z \neq z'$.
Let 
\begin{equation*}
    n = \min m \in \mathbb{N} \such z_m \neq z'_m
\end{equation*}
We have
\begin{equation*}
    \forall m < n, \; z_m = z'_m
\end{equation*}
we have $z_n, z'_n \in X$, but $z_n \neq z'_n$.
Let us assume without loss of generality that $z_n < z'_n$.
Because $z$ is strictly increasing we have
\begin{equation*}
    \forall m < n, \; \rp{z'_m = z_m} \neq z_n
\end{equation*}
and because $z'$ is strictly increasing we have
\begin{equation*}
    \forall m > n, \; z'_m > z'_n > z_n
\end{equation*}

In summary, we have
\begin{gather*}
    \forall m < n, \; z'_m \neq z_n \\
    z'_n \neq z_n \\
    \forall m > n, \; z'_m \neq z_n
\end{gather*}
Which proves that $z_n \in X$ does not appear anywhere in $z'$, contradicts the hypothesis, and proves the lemma.
\end{proof}
\end{lemma}

\subsection{Correctness}
\label{section:tob.correctness}

\Cref{subsection:tob.noduplicationintegrity} proves No Duplication and Integrity.
\Cref{subsection:tob.agreementtotalorder} proves Agreement and Total Order.
\Cref{subsection:tob.validity} proves Validity.

\subsubsection{No Duplication and Integrity}
\label{subsection:tob.noduplicationintegrity}

\begin{observation}
\label{observation:tob.deliveredgrows}
Let $\server$ be a correct server.
Upon initialization, $\server$ sets $delivered = \cp{}$ (line \ref{line:tob.initdelivered}). 
$\server$ never removes elements from $delivered$. 
This means $delivered$ is non-decreasing at $\server$.
Equivalently, $\server$ satisfies $\rp{\client, m} \in delivered$ if and only if $\server$ added $\rp{\client, m}$ to $delivered$.
\end{observation}

\begin{observation}
\label{observation:tob.proposedgrows}
Let $\server$ be a correct server.
Upon initialization, $\server$ sets $proposed = \cp{}$ (line \ref{line:tob.initproposed}). 
$\server$ never removes elements from $proposed$. 
This means $proposed$ is non-decreasing at $\server$.
Equivalently, $\server$ satisfies $\rp{\client, m, b} \in proposed$ if and only if $\server$ added $\rp{\client, m, b}$ to $proposed$.
\end{observation}

\begin{theorem}
\label{theorem:tob.noduplication}
\tobal satisfies No Duplication.
\begin{proof}
Let $\server$ be a correct server.
$\server$ triggers $\event{\tobin}{Deliver}{\client, m}$ (line \ref{line:tob.deliver} only) only if $\rp{\client, m} \notin delivered$ (line \ref{line:tob.delivercheck}).
Immediately after triggering $\event{\tobin}{Deliver}{\client, m}$, $\server$ adds $\rp{\client, m}$ to $delivered$ (line \ref{line:tob.adddelivered}).
The theorem follows immediately from Observation \ref{observation:tob.deliveredgrows}
\end{proof}
\end{theorem}

\begin{lemma}
\label{lemma:tob.clientsubmissions}
Let $\client$ be a correct client.
For every $m \in submissions$ at $\client$, $\client$ triggered $\event{\tobin}{Broadcast}{m}$.
\begin{proof}
Upon initialization, $\client$ sets $submissions = \cp{}$ (line \ref{line:tobc.initsubmissions}).
$\client$ sets $submissions\qp{m}$ (line \ref{line:tobc.setsubmission} only) only upon executing $submit\rp{m, \any}$ (line \ref{line:tobc.submit}).
$\client$ invokes $submit\rp{m, \any}$ only from lines \ref{line:tobc.submituponbroadcast} and \ref{line:tobc.submituponfailure}, respectively: upon triggering $\event{\tobin}{Broadcast}{m}$ (line \ref{line:tobc.broadcast}); and only if $m \in submissions$ (line \ref{line:tobc.failurecondition}).
By induction, for every $m \in submissions$ at $\server$, $\server$ triggered $\event{\tobin}{Broadcast}{m}$.
\end{proof}
\end{lemma}

\begin{lemma}
\label{lemma:tob.clientmessages}
Let $\client$ be a correct client, let $m \in \messages$.
If $\client$ issues any $\qp{\texttt{Message}, m, \any}$ message, then $\client$ triggered $\event{\tobin}{Broadcast}{m}$.
\begin{proof}
Immediately after issuing a $\qp{\texttt{Message}, m, \any}$ message (line \ref{line:tobc.sendmessage} only), $\client$ sets $submissions\qp{m}$ (line \ref{line:tobc.setsubmission}).
The lemma immediately follows from Lemma \ref{lemma:tob.clientsubmissions}.
\end{proof}
\end{lemma}

\begin{theorem}
\label{theorem:tob.integrity}
\tobal satisfies Integrity.
\begin{proof}
Let $\server$ be a correct server, let $\client$ be a correct client, let $m \in \messages$ such that $\server$ triggers $\event{\tobin}{Deliver}{\client, m}$.
$\server$ triggers $\event{\tobin}{Deliver}{\client, m}$ (line \ref{line:tob.deliver} only) only upon executing $order\rp{\client, m, b}$ for some $b \in \altime$ (line \ref{line:tob.order}).
$\server$ invokes $order\rp{\client, m, b}$ (line \ref{line:tob.invokeorder} only) only if $decisions\qp{\rp{\client, m, b}} = \true$ (line \ref{line:tob.ordercondition}).
$\server$ sets $decisions\qp{\rp{\client, m, b}} = \true$ (line \ref{line:tob.setdecision} only) only upon triggering $\event{\conin\qp{\rp{\client, m, b}}}{Decide}{\true}$ (line \ref{line:tob.decide}).
By the \validity property of $\conin\qp{\rp{\client, m, b}}$, at least one correct server $\serversecondary$ triggered $\event{\conin\qp{\client, m, b}}{Propose}{\true}$.
$\serversecondary$ triggers $\event{\conin\qp{\rp{\client, m, b}}}{Propose}{\true}$ (line \ref{line:tob.proposeuponmessage} only) only upon delivering a $\qp{\texttt{Message}, m, b}$ from $\client$.
By Lemma \ref{lemma:tob.clientmessages}, $\client$ triggered $\event{\tobin}{Broadcast}{m}$, and the theorem is proved.
\end{proof}
\end{theorem}

\subsubsection{Agreement and Total Order}
\label{subsection:tob.agreementtotalorder}

\begin{assumption}
\label{assumption:tob.ordering}
$\rp{\clients \times \messages \times \altime}$ is ordered by successively comparing the time, client, and message components:
\begin{equation*}
    \biggl(\rp{\client, m, b} \geq \rp{\client', m', b'}\biggr) \stackrel{def}{\Longleftrightarrow} \biggl(\rp{b > b'} \vee \rp{b = b' \wedge \client > \client'} \vee \rp{b = b' \wedge \client = \client' \wedge m \geq m'}\biggr)
\end{equation*}
The ordering criterion chosen for clients and messages is not important.
Adequate criteria include, \eg lexicographic ordering of client identifiers and message payloads.
\end{assumption}

\begin{observation}
\label{observation:tob.observedgrows}
Let $\server$ be a correct server.
Upon initialization, $\server$ sets $observed = \cp{}$ (line \ref{line:tob.initobserved}). 
$\server$ never removes elements from $observed$. 
This means $observed$ is non-decreasing at $\server$.
Equivalently, $\server$ satisfies $\rp{\client, m, b} \in observed$ if and only if $\server$ added $\rp{\client, m, b}$ to $observed$.
\end{observation}

\begin{definition}[Universe]
\label{definition:tob.universe}
The \textbf{universe} $\universe \subseteq \rp{\clients \times \messages \times \altime}$ includes $\rp{\client, m, b}$ if and only if any correct server eventually adds $\rp{\client, m, b}$ to $observed$.
\end{definition}

\begin{notation}[Universe, time-cropped]
For all $t \in \altime$, we use
\begin{equation*}
    \universe[\leq t] = \cp{\rp{\client, m, b} \in \universe \such b \leq t}
\end{equation*}
\end{notation}

\begin{lemma}
\label{lemma:tob.spot}
Let $\server$ be a correct server, let $\rp{\client, m, b} \in \rp{\clients, \messages, \altime}$.
Upon returning from an invocation to $spot\rp{\client, m, b}$, $\server$ satisfies $\rp{\client, m, b} \in observed$.
\begin{proof}
Upon invoking $spot\rp{\client, m, b}$, $\server$ either satisfies $\rp{\client, m, b} \in observed$ or $\rp{\client, m, b} \notin observed$.
If $\rp{\client, m, b} \in observed$, the lemma holds trivially.
If $\rp{\client, m, b} \notin observed$ (line \ref{line:tob.spotcondition}) then $\server$ adds $\rp{\client, m, b}$ to $observed$ (line \ref{line:tob.addobserved}).
\end{proof}
\end{lemma}

\begin{lemma}
\label{lemma:tob.observedtotality}
Let $\rp{\client, m, b} \in \universe$. 
Eventually, every correct server adds $\rp{\client, m, b}$ to $observed$.
\begin{proof}
By Definition \ref{definition:tob.universe}, some correct server $\server$ eventually adds $\rp{\client, m, b}$ to $observed$.
$\server$ does so only by executing line \ref{line:tob.addobserved}.
Immediately before doing so, $\server$ sends an $\qp{\texttt{Observe}, \client, m, b}$ message to every server (lines \ref{line:tob.sendobservefor} and \ref{line:tob.sendobserve}).
Upon delivering $\server$'s $\qp{\texttt{Observe}, \client, m, b}$ message (line \ref{line:tob.deliverobserve}), a correct server $\serversecondary$ invokes $spot\rp{\client, m, b}$ (line \ref{line:tob.spotuponobserve}).
The lemma immediately follows from Lemma \ref{lemma:tob.spot} and Observation \ref{observation:tob.observedgrows}.
\end{proof}
\end{lemma}

\begin{lemma}
\label{lemma:tob.decisiontotality}
Let $\rp{\client, m, b} \in \universe$. 
Eventually, every correct server triggers $\event{\conin\qp{\rp{\client, m, b}}}{Decide}{}$.
\begin{proof}
By Lemma \ref{lemma:tob.observedtotality}, every correct server eventually adds $\rp{\client, m, b}$ to $observed$.
Let $\server$ be a correct server.
Let $a$ denote the time when $\server$ adds $\rp{\client, m, b}$ to $observed$.
By Observation \ref{observation:tob.observedgrows}, $\server$ satisfies $\rp{\client, m, b} \in observed$ forever after $a$.

Upon initialization, $\server$ sets $proposed = \cp{}$ (line \ref{line:tob.initproposed}).
Moreover, $\server$ adds $\rp{\client, m, b}$ to $proposed$ (lines \ref{line:tob.addproposeduponmessage} and \ref{line:tob.addproposeduponexpire}) only after triggering $\event{\conin\qp{\rp{\client, m, b}}}{Propose}{}$ (lines \ref{line:tob.proposeuponmessage} and \ref{line:tob.proposeuponexpire}).
Finally, $\server$ never triggers $\event{\conin\qp{\rp{\client, m, b}}}{Propose}{}$ (lines \ref{line:tob.proposeuponmessage} and \ref{line:tob.proposeuponexpire}) unless $\rp{\client, m, b} \notin proposed$ (lines \ref{line:tob.proposedcheckuponmessage} and \ref{line:tob.proposeexpirecondition}).
In summary, if $\rp{\client, m, b} \in proposed$ at $\server$, then $\server$ triggered $\event{\conin\qp{\rp{\client, m, b}}}{Propose}{}$ exactly once.

At all times after $\max\rp{a, b}$, $\server$ satisfies $\rp{\client, m, b} \in observed$ and $b < local\_time\rp{}$. 
As a result, $\server$ eventually adds $\rp{\client, m, b}$ to $proposed$.
Indeed, let us assume, for the sake of contradiction, that $\server$ never adds $\rp{\client, m, b}$ to $proposed$.
$\server$ would forever satisfy the condition at line \ref{line:tob.proposeexpirecondition}.
This would result in $\server$ eventually triggering the condition, thus adding $\rp{\client, m, b}$ to $proposed$ (line \ref{line:tob.addproposeduponexpire}).

In summary, $\server$ eventually adds $\rp{\client, m, b}$ to $proposed$.
As a result, $\server$ triggers $\event{\conin\qp{\rp{\client, m, b}}}{Propose}{}$ once.
By the Termination property of $\conin\qp{\rp{\client, m, b}}$, $\server$ eventually triggers $\event{\conin\qp{\rp{\client, m, b}}}{Decide}{}$, and the lemma is proved.
\end{proof}
\end{lemma}

\begin{definition}[Filtered universe]
\label{definition:tob.filtereduniverse}
The \textbf{filtered universe} $\filtered \subseteq \universe$ includes $\rp{\client, m, b}$ if and only if the decision of $\conin\qp{\rp{\client, m, b}}$ is $\true$.
\end{definition}

We underline that, by Lemma \ref{lemma:tob.decisiontotality} and the Agreement property of Consensus, Definition \ref{definition:tob.filtereduniverse} is well-formed.

\begin{notation}[Filtered universe, time-cropped]
\label{notation:filtereduniversetimecropped}
For all $t \in \altime$, we use
\begin{equation*}
    \filtered[\leq t] = \filtered \cap \universe[\leq t]
\end{equation*}
\end{notation}

\begin{observation}
\label{observation:remotetimesgrow}
Let $\server$ be a correct server, let $\serversecondary$ be a server.
Upon initialization, $\server$ sets $remote\_times\qp{\serversecondary} = -\infty$ (line \ref{line:tob.initremotetimes}).
$\server$ updates $remote\_times\qp{\serversecondary}$ only to a value greater or equal to $remote\_times\qp{\serversecondary}$ (line \ref{line:tob.updateremotetimes}).
This means that $remote\_times\qp{\serversecondary}$ is non-decreasing at $\server$.
\end{observation}

\begin{observation}
\label{observation:tob.locktimegrows}
Let $\server$ be a correct server.
$lock\_time\rp{}$ (line \ref{line:tob.locktime}) returns the largest $t$ such that at least $4\faultcount + 1$ values in $remote\_times$ are greater or equal to $t$.
By Observation \ref{observation:remotetimesgrow}, this means that $lock\_time\rp{}$ is non-decreasing at $\server$.
\end{observation}

\begin{lemma}
\label{lemma:tob.locktimegrowsindefinitely}
Let $\server$ be a correct server, let $t \in \altime$.
If $\abs{\universe} = \infty$, then $\server$ eventually satisfies $lock\_time\rp{} \geq t$.
\begin{proof}
We start by noting that a correct server adds an element to $observed$ (line \ref{line:tob.addobserved}) only after scheduling $beat\rp{}$ for execution (line \ref{line:tob.schedulebeat}).
By Lemma \ref{lemma:tob.observedtotality}, every correct server invokes $beat\rp{}$ infinite times.
Let $t^*$ denote any moment when every correct server satisfies $local\_time\rp{} \geq t$.
Eventually, every correct server $\serversecondary$ invokes $beat\rp{}$ after $t^*$.
Upon doing so, $\serversecondary$ sends a 
\begin{equation*}
\qp{\texttt{Time}, \rp{t' = local\_time\rp{}} \geq t}
\end{equation*}
message to $\server$ (lines \ref{line:tob.sendtimefor} and \ref{line:tob.sendtime}).
Upon delivering $\serversecondary$'s $\qp{\texttt{Time}, t'}$ message (line \ref{line:tob.delivertime}) $\server$ sets $remote\_times\qp{\serversecondary} \geq t'$ (line \ref{line:tob.updateremotetimes}). 
By Observation \ref{observation:remotetimesgrow}, $\server$ eventually satisfies $remote\_times\qp{\serversecondary} \geq t$ for every correct server $\serversecondary$.
As we assume at most $f$ faulty servers, $\server$ eventually has $4 \faultcount + 1$ values in $remote\_times$ that are greater or equal to $t$.
$lock\_time\rp{}$ (line \ref{line:tob.locktime}) returns the highest $\hat t$ such that at least $4 \faultcount + 1$ values in $remote\_times$ are greater or equal to $t$ (line \ref{line:tob.locktimereturn}).
Eventually, we have $\hat t \geq t$, and the lemma is proved.
\end{proof}
\end{lemma}

\begin{observation}
\label{observation:tob.candidatesgrows}
Let $\server$ be a correct server.
Upon initialization, $\server$ sets $candidates = \cp{}$ (line \ref{line:tob.initcandidates}). 
$\server$ never removes elements from $candidates$. 
This means $candidates$ is non-decreasing at $\server$.
Equivalently, $\server$ satisfies $\rp{\client, m, b} \in candidates$ if and only if $\server$ added $\rp{\client, m, b}$ to $candidates$.
\end{observation}

\begin{definition}[Candidates]
\label{definition:tob.candidates}
Let $\server$ be a correct server.
$\server$'s \textbf{candidates} $\candidates{\server} \subseteq \rp{\clients \times \messages \times \altime}$ includes $\rp{\client, m, b}$ if and only if $\server$ eventually adds $\rp{\client, m, b}$ to $candidates$.
\end{definition}

\begin{lemma}
\label{lemma:tob.candidatesinuniverse}
Let $\server$ be a correct server.
We have $\candidates{\server} \subseteq \universe$.
\begin{proof}
Let $\rp{\client, m, b} \in \candidates{\server}$.
$\server$ adds $\rp{\client, m, b}$ to $candidates$ (line \ref{line:tob.addcandidate} only) only upon executing $spot\rp{\client, m, b}$.
By Lemma \ref{lemma:tob.spot} and Definition \ref{definition:tob.universe} we then have $\rp{\client, m, b} \in \universe$, and the lemma is proved.
\end{proof}
\end{lemma}

\begin{notation}[Candidates, time-cropped]
\label{notation:candidatestimecropped}
Let $\server$ be a correct server. For all $t \in \altime$, we use
\begin{equation*}
    \candidates{\server}[\leq t] = \cp{\rp{\client, m, b} \in \candidates{\server} \such b \leq t}
\end{equation*}
\end{notation}

\begin{lemma}
\label{lemma:tob.croppedcandidatesarefinite}
Let $\server$ be a correct server, let $t \in \altime$. We have $\abs{\candidates{\server}[\leq t]} < \infty$.
\begin{proof}
We start by noting that $\server$ adds $\rp{\client, m, b}$ to $candidates$ (line \ref{line:tob.addcandidate} only) only upon invoking $spot\rp{\client, m, b}$.
By Lemma \ref{lemma:tob.spot}, at any point in time $\server$ must thereofore satisfy
\begin{equation*}
    \abs{candidates} \leq \abs{observed} + 1
\end{equation*}
(the equality can only hold, briefly, whenever $\server$ has added some $\rp{\client, m, b}$ to $candidates$ (line \ref{line:tob.addcandidate}) but not yet to $observed$ (line \ref{line:tob.addobserved})).
As a result, if $\abs{\universe} < \infty$, by Definition \ref{definition:tob.universe} $\server$ always satisfies
\begin{equation*}
    \abs{candidates} \leq \abs{observed} + 1 \leq \abs{\universe} + 1 < \infty
\end{equation*}
and the lemma is immediately proved. 
Throughout the remainder of this proof, we assume $\abs{\universe} = \infty$.

Let $t^*$ denote any time when $\server$ satisfies $lock\_time\rp{} > t$.
By Lemma \ref{lemma:tob.locktimegrowsindefinitely}, $t^*$ is guaranteed to exist.
$\server$ invokes $spot$ (lines \ref{line:tob.spotuponmessage} and \ref{line:tob.spotuponobserve} only) only upon delivering a $\texttt{Message}$ (line \ref{line:tob.delivermessage}) or $\texttt{Observe}$ (line \ref{line:tob.spotuponobserve}) message.
Moreover, $\server$ adds $\rp{\client, m, b \leq t}$ to $candidates$ (line \ref{line:tob.addcandidate} only) only if $\server$ satisfies $lock\_time\rp{} < \rp{t \geq b}$ (line \ref{line:tob.candidatecondition}).
By the two above, Observation \ref{observation:remotetimesgrow} and Notation \ref{notation:candidatestimecropped}, we then have that $\abs{\candidates{\server}[\leq t]}$ is bounded by the number of messages $\server$ delivers by time $t^*$.
As $\server$ cannot deliver infinite messages in finite time, we have $\abs{\candidates{\server}[\leq t]} < \infty$, and the lemma is proved.
\end{proof}
\end{lemma}

\begin{lemma}
\label{lemma:tob.observebeforetime}
Let $\rp{\client, m, b} \in \universe$ such that some correct server $\server$ triggers $\event{\conin\qp{\rp{\client, m, b}}}{Propose}{\true}$. $\server$ sends an $\qp{\texttt{Observe}, \client, m, b}$ message to every server before sending any $\qp{\texttt{Time}, t \geq b}$ to any server.
\begin{proof}
We start by noting that, by Observation \ref{observation:tob.observedgrows}, if $\server$ satisfies $\rp{\client, m, b} \in observed$, then $\server$ previously sent an $\qp{\texttt{Observe}, \client, m, b}$ message to every server.
Indeed, $\server$ adds $\rp{\client, m, b}$ to $observed$ (line \ref{line:tob.addobserved} only) only after sending an $\qp{\texttt{Observe}, \client, m, b}$ message to every server (lines \ref{line:tob.sendobservefor} and \ref{line:tob.sendobserve}).

$\server$ triggers $\event{\conin\qp{\rp{\client, m, b}}}{Propose}{\true}$ (line \ref{line:tob.proposeuponmessage} only) only if, upon evaluating line \ref{line:tob.intime} at some time $t^*$, $\server$ finds that $local\_time\rp{} < b$.
Before evaluating line \ref{line:tob.intime}, $\server$ invokes $spot\rp{\client, m, b}$ (line \ref{line:tob.spotuponmessage}).
Consequently, by Lemma \ref{lemma:tob.spot}, by time $t^*$ $\server$ sent an $\qp{\texttt{Observe}, \client, m, b}$ message to every server.

$\server$ issues a $\qp{\texttt{Time}, t \geq b}$ (line \ref{line:tob.sendtime} only) only if $\server$ satisfies $local\_time\rp{} \geq b$.
Because a correct server's local time is non-decreasing, if $\server$ issues any $\qp{\texttt{Time}, t \geq b}$ message, $\server$ does so after $t^*$: the lemma is proved.
\end{proof}
\end{lemma}

\begin{lemma}
\label{lemma:tob.onlycandidatesarefiltered}
Let $\server$ be a correct server. We have $\filtered \subseteq \candidates{\server}$.
\begin{proof}
Let $\rp{\client, m, b} \in \filtered$.
Let $\serverssubset$ denote the set of correct servers that trigger $\event{\conin\qp{\rp{\client, m, b}}}{Propose}{\true}$.
By Definition \ref{definition:tob.filtereduniverse} and the \validity property of $\conin\qp{\rp{\client, m, b}}$, we have $\abs{\serverssubset} \geq \faultcount + 1$.
As an immediate consequence, we have $\rp{\client, m, b} \in \universe$.
Indeed, let $\serversecondary \in \serverssubset$.
Immediately before triggering $\event{\conin\qp{\rp{\client, m, b}}}{Propose}{\true}$ (line \ref{line:tob.proposeuponmessage} only), $\serversecondary$ invokes $spot\rp{\client, m, b}$ (line \ref{line:tob.spotuponmessage}): by Lemma \ref{lemma:tob.spot}, $\rp{\client, m, b} \in \universe$.

The lemma trivially holds true if $\server$ never satisfies $lock\_time\rp{} \geq b$.
In that case, before adding $\rp{\client, m, b}$ to $observed$ (line \ref{line:tob.addobserved} only), $\server$ finds that $b > lock\_time\rp{}$ (line \ref{line:tob.candidatecondition}) and adds $\rp{\client, m, b}$ to $candidates$ (line \ref{line:tob.addcandidate}).
Throughout the remainder of this proof, we assume that server eventually satisfies $lock\_time\rp{} \geq b$.

Let $t^+$ denote the moment when $\server$ first satisfies $lock\_time\rp{} \geq b$: $\server$ satisfies $lock\_time\rp{} < b$ at all times before $t^+$.
In order for $\server$ to satisfy $lock\_time\rp{} \geq b$ (line \ref{line:tob.locktime}), $remote\_times$ must contain at least $4 \faultcount + 1$ values that are greater or equal to $b$.
This means that the set $\serverssubsetsecondary \subseteq \servers$ that includes $\servertertiary$ if and only if $remote\_times\qp{\servertertiary} > b$ at $\server$ at time $t^+$ contains at least $4 \faultcount + 1$ elements.

By the pigeonhole principle, $\serverssubset$ and $\serverssubsetsecondary$ intersect in at least one correct server $\serversecondary$.
$\server$ sets $remote\_times\qp{\serversecondary} \geq b$ (line \ref{line:tob.updateremotetimes} only) only upon delivering a $\qp{\texttt{Time}, t \geq b}$ message from $\serversecondary$ (line \ref{line:tob.delivertime}).
This means that, by time $t^+$, $\server$ delivered a $\qp{\texttt{Time}, t \geq b}$ message from $\serversecondary$.
By Lemma \ref{lemma:tob.observebeforetime} and the FIFO property of FIFO links, $\server$ does not deliver any $\qp{\texttt{Time}, t \geq b}$ message from $\serversecondary$ before delivering $\serversecondary$'s $\qp{\texttt{Observe}, \client, m, b}$ message.
In summary, $\server$ delivers a $\qp{\texttt{Observe}, \client, m, b}$ message before $t^*$.
Upon doing so (line \ref{line:tob.deliverobserve}), $\server$ invokes $spot\rp{\client, m, b}$ (line \ref{line:tob.spotuponobserve}), finds that $lock\_time\rp{} < b$ (line \ref{line:tob.candidatecondition}) and adds $\rp{\client, m, b}$ to $candidates$ (line \ref{line:tob.addcandidate}).
This proves $\rp{\client, m, b} \in \candidates{\server}$ and concludes the lemma.
\end{proof}
\end{lemma}

\begin{corollary}
\label{corollary:tob.croppedfilteredisfinite}
For all $t \in \altime$ we have $\abs{\filtered[\leq t]} < \infty$.
\begin{proof}
By Definitions \ref{definition:tob.filtereduniverse} and \ref{definition:tob.candidates} and Lemmas \ref{lemma:tob.croppedcandidatesarefinite} and \ref{lemma:tob.onlycandidatesarefiltered}, for any correct server $\server$ we have
\begin{equation*}
    \abs{\rp{\filtered[\leq t] \subseteq \candidates{\server}[\leq t]}} < \infty
\end{equation*}
\end{proof}
\end{corollary}

\begin{corollary}
\label{corollary:tob.filteredenumeration}
One and only one strictly increasing sequence $\filtered$ exists such that
\begin{equation*}
    \forall \rp{\client, m, b} \in \filtered, \; \exists n \such \filtered_n = \rp{\client, m, b}
\end{equation*}
\begin{proof}
It follows immediately from Corollary \ref{corollary:tob.croppedfilteredisfinite} and Lemmas \ref{lemma:tob.enumerateset} and \ref{lemma:tob.enumerationsareunique}.
\end{proof}
\end{corollary}

\begin{notation}[Filtered universe as a sequence]
We use $\filtered$ to interchangeably denote the filtered universe as a set (Definition \ref{definition:tob.filtereduniverse}) and as a strictly increasing enumeration (Corollary \ref{corollary:tob.filteredenumeration}).
\end{notation}

\begin{definition}[Order calls]
\label{definition:tob.order}
Let $\server$ be a correct server. 
$\server$'s sequence of \textbf{order calls} 
\begin{equation*}
    \order{\server} \in \rp{\clients \times \messages \times \altime}^{\leq \infty}
\end{equation*}
captures the sequence of invocations $\server$ makes to the $order$ procedure (line \ref{line:tob.order}).
\end{definition}

For example, if $\server$ invokes $order$ exactly three times, $order\rp{\client, m, b}$ first, $order\rp{\client', m', b'}$ second and $order\rp{\client'', m'', b''}$ third, then
\begin{equation*}
    \order{\server} = \qp{\rp{\client, m, b}, \rp{\client', m', b'}, \rp{\client'', m'', b''}}
\end{equation*}

\begin{lemma}
\label{lemma:tob.orderincreases}
Let $\server$ be a correct server. $\order{\server}$ is strictly increasing.
\begin{proof}
Let $k, n \leq \abs{\order{\server}}$ such that $k < n$.
Let $\rp{\client, m, b} = \order{\server}_k$, let $\rp{\client', m', b'} = \order{\server}_n$.
Immediately after invoking $order\rp{\client, m, b}$ (line \ref{line:tob.invokeorder}), $\server$ sets $last\_processed = \rp{\client, m, b}$ (line \ref{line:tob.updatelastprocessed}). 
By Observation \ref{observation:tob.lastprocessedgrows}, when $\server$ invokes $order\rp{\client', m', b'}$  we still have $last\_order \geq \rp{\client, m, b}$.
$\server$ invokes $order\rp{\client', m', b'}$ (line \ref{line:tob.invokeorder}), however, only if $\rp{\client', m', b'} > last\_processed$ (line \ref{line:tob.processcondition}).
To summarize, when $\server$ invokes $order\rp{\client', m', b'}$, we have
\begin{equation*}
    \rp{\client', m', b'} > last\_processed \geq \rp{\client, m, b}
\end{equation*}
which proves $\order{\server}_k < \order{\server}_n$ and concludes the lemma.
\end{proof}
\end{lemma}

\begin{lemma}
\label{lemma:tob.orderedarefiltered}
Let $\server$ be a correct server, let $n \leq \abs{\order{\server}}$.
We have $\order{\server}_n \in \filtered$.

\begin{proof}
$\server$ invokes $order\rp{\client, m, b}$ (line \ref{line:tob.invokeorder} only) only if $decisions\qp{\rp{\client, m, b}} = \true$ (line \ref{line:tob.ordercondition}).
Upon initialization, $\server$ sets $decisions = \cp{}$ (line \ref{line:tob.initdecisions}). 
$\server$ sets $decisions\qp{\rp{\client, m, b}} = \true$ (line \ref{line:tob.setdecision} only) only upon triggering $\event{\conin\qp{\rp{\client, m, b}}}{Decide}{\true}$ (line \ref{line:tob.decide}). 
The lemma follows immediately from Definition \ref{definition:tob.filtereduniverse}.
\end{proof}
\end{lemma}

\begin{observation}
\label{observation:tob.lastprocessedgrows}
Let $\server$ be a correct server.
$\server$ updates $last\_processed$ to $\rp{\client, m, b}$ (line \ref{line:tob.updatelastprocessed} only) only if $\rp{\client, m, b} > last\_processed$ (line \ref{line:tob.processcondition}).
This means that $last\_processed$ is non-decreasing at $\server$.
\end{observation}

\begin{lemma}
\label{lemma:tob.newcandidatebeyondlastprocessed}
Let $\server$ be a correct server, let $\rp{\client, m, b} \in \candidates{\server}$.
Upon adding $\rp{\client, m, b}$ to $candidates$, $\server$ satisfies $\rp{\client, m, b} > last\_delivered$.
\begin{proof}
Let $\rp{\client^*, m^*, b^*}$ denote the value of $last\_delivered$ at $\server$ when $\server$ adds $\rp{\client, m, b}$ to $candidates$.
The lemma trivially holds if $b^* = -\infty$ (line \ref{line:tob.initlastprocessed}).
Throughout the remainder of this proof, we assume $b^* > -\infty$.
Upon setting $last\_processed = \rp{\client^*, m^*, b^*}$ (line \ref{line:tob.updatelastprocessed} only), $\server$ satisfied $b^* \leq lock\_time\rp{}$ (line \ref{line:tob.processcondition}).
When $\server$ adds $\rp{\client, m, b}$ to $candidates$ (line \ref{line:tob.addcandidate} only), however, $\server$ satisfies $b > lock\_time\rp{}$ (line \ref{line:tob.candidatecondition}).
By Observation \ref{observation:tob.locktimegrows}, we then have $b > b^*$, and the lemma immediately follows from Assumption \ref{assumption:tob.ordering}.
\end{proof}
\end{lemma}

\begin{lemma}
\label{lemma:tob.lastprocessedovertake}
Let $\server$ be a correct server. 
Let $\rp{\client, m, b} \in \candidates{\server}$.
$\server$ does not satisfy $last\_processed \geq \rp{\client, m, b}$, before invoking $order\rp{\client, m, b}$.

\begin{proof}
Let us assume $\server$ eventually satisfies $last\_processed \geq \rp{\client, m, b}$. 
Let $\rp{\client^+, m^+, b^+}$ denote the first value $last\_processed$ takes at $\server$ such that $\rp{\client^+, m^+, b^+} \geq \rp{\client, m, b}$.
Let $\rp{\client^-, m^-, b^-}$ denote the value $last\_processed$ had immediately before being updated to $\rp{\client^+, m^+, b^+}$.
We have
\begin{equation*}
    \rp{\client^-, m^-, b^-} < \rp{\client, m, b} \leq \rp{\client^+, m^+, b^+}
\end{equation*}
Noting that $\server$ initializes $last\_processed$ to $\rp{\bot, \bot, -\infty < b}$ (line \ref{line:tob.initlastprocessed}), $\server$ can set $last\_processed = \rp{\client^+, m^+, b^+}$ (line \ref{line:tob.updatelastprocessed} only) only if
\begin{equation}
\label{equation:tob.filteredareordered.processcondition}
    \rp{\client^+, m^+, b^+} = \min \cp{\rp{\client', m', b'} \in candidates \such \rp{\client', m', b'} > \rp{\client^-, m^-, b^-}}
\end{equation}
(line \ref{line:tob.processcondition}).

By Definition \ref{definition:tob.candidates}, $\server$ eventually adds $\rp{\client, m, b}$ to $candidates$.
By Lemma \ref{lemma:tob.newcandidatebeyondlastprocessed}, when $\server$ does so, $\server$ satisfies $last\_processed < \rp{\client, m, b}$.
As such, $\server$ satisfies $\rp{\client, m, b} \in candidates$ before setting $last\_processed = \rp{\client^+, m^+, b^+}$, $\server$.
By Equation \ref{equation:tob.filteredareordered.processcondition}, this proves $\rp{\client^+, m^+, b^+} \leq \rp{\client, m, b}$.
In summary, we have
\begin{align*}
    \rp{\client^+, m^+, b^+} &\geq \rp{\client, m, b} \\
    \rp{\client^+, m^+, b^+} &\leq \rp{\client, m, b}
\end{align*}
Which proves that $\server$ eventually sets $last\_processed = \rp{\client, m, b}$.
$\server$ does so (line \ref{line:tob.processcondition}) only when $\rp{\client, m, b} \in decisions$.
Because $\rp{\client, m, b} \in \filtered$, we must then have $decisions\qp{\rp{\client, m, b}} = \true$.
This means that, immediately before setting $last\_processed = \rp{\client, m, b}$, $\server$ satisfies the condition at line \ref{line:tob.ordercondition}.
As a result, $\server$ invokes $order\rp{\client, m, b}$ (line \ref{line:tob.invokeorder}), and the result is proved.
\end{proof}
\end{lemma}

\begin{lemma}
\label{lemma:tob.filteredareordered}
Let $\server$ be a correct server. Let $\rp{\client, m, b} \in \filtered$. For some $n$, we have $\order{\server}_n = \rp{\client, m, b}$.
\begin{proof}
By Definition \ref{definition:tob.order}, the lemma reduces to proving that $\server$ eventually invokes $order\rp{\client, m, b}$.

We start by noting that, unless $last\_processed = \rp{\bot, \bot, -\infty}$ (line \ref{line:tob.initlastprocessed}), we have $last\_processed \in \candidates{\server}$.
Indeed, $\server$ sets $last\_processed = \rp{\client', m', b'}$ (line \ref{line:tob.updatelastprocessed} only) only if $\rp{\client', m', b'} \in candidates$ (line \ref{line:tob.processcondition}).
Moreover, by Observation \ref{observation:tob.lastprocessedgrows}, $last\_processed$ is non-decreasing.
Finally, because by Lemma \ref{lemma:tob.croppedcandidatesarefinite} we have $\abs{\candidates{\server}{\leq b}} < \infty$, $\server$ can update the value of $last\_processed$ only a finite number of times before $last\_processed \geq \rp{\client, m, b}$.
Every time $\server$ triggers the condition at line \ref{line:tob.processcondition}, $last\_processed$ is updated to a new value (line \ref{line:tob.updatelastprocessed}).
At any point in time, let
\begin{equation*}
    \rp{\client', m', b'} = \min \rp{\hat \client, \hat m, \hat b} \in candidates \such \rp{\hat \client, \hat m, \hat b} > last\_processed
\end{equation*}
(line \ref{line:tob.processcondition}). 
As long as $last\_processed < \rp{\client, m, b}$, $\rp{\client', m', b'}$ is guaranteed to exist.
By Lemma \ref{lemma:tob.decisiontotality}, $\server$ eventually satisfies $\rp{\client', m', b'} \in decisions$.
By Lemma \ref{lemma:tob.locktimegrowsindefinitely}, $\server$ eventually satisfies $b' \leq lock\_time\rp{}$.
This means that the condition at line \ref{line:tob.processcondition} is eventually guaranteed to trigger, and $\server$ will update $last\_processed$ to a new value.
In summary, $\server$ keeps updating $last\_processed$ to a new value as long as $last\_processed < \rp{\client, m, b}$.
Having done so a finite number of times, $\server$ eventually satisfies $last\_processed \geq \rp{\client, m, b}$.
Moreover, by Lemma \ref{lemma:tob.onlycandidatesarefiltered}, we have $\rp{\client, m, b} \in \candidates{\server}$.
As a result, by Lemma \ref{lemma:tob.lastprocessedovertake}, $\server$ eventually invokes $order\rp{\client, m, b}$, and the lemma is proved.
\end{proof}
\end{lemma}

\begin{corollary}
\label{corollary:orderequalfiltered}
Let $\server$ be a correct server.
We have $\order{\server} = \filtered$.

\begin{proof}
It follows immediately from Lemmas \ref{lemma:tob.orderincreases}, \ref{lemma:tob.orderedarefiltered}, \ref{lemma:tob.filteredareordered} and \ref{lemma:tob.enumerationsareunique}. 
\end{proof}
\end{corollary}

\begin{corollary}
\label{corollary:tob.ordersmatch}
Let $\server, \server'$ be correct servers.
We have $\order{\server} = \order{\server'}$.
\begin{proof}
It follows immediately from Corollary \ref{corollary:orderequalfiltered}.
\end{proof}
\end{corollary}

\begin{theorem}
\label{theorem:tob.agreementtotalorder}
\tobal satisfies Agreement and Total Order.
\begin{proof}
A correct server triggers $\event{\tobin}{Deliver}{}$ (line \ref{line:tob.deliver} only) only upon executing $order$ (line \ref{line:tob.order}).
Procedure $order$ implements a deterministic state machine, using only $delivered$ as state: $order$ reads from / writes to only $delivered$, $delivered$ is read from / written to only by $order$.
Moreover, every correct server initializes $delivered$ to the same value (line \ref{line:tob.initdelivered}).
Finally, by Corollary \ref{corollary:tob.ordersmatch}, every correct server issues the same sequence of invocations to $order$.
As a result, every correct server triggers the same sequence of $\event{\tobin}{Deliver}{}$ events, and the theorem is proved.
\end{proof}
\end{theorem}

\subsubsection{Validity}
\label{subsection:tob.validity}

\begin{lemma}
\label{lemma:tob.orderedaredelivered}
Let $\server$ be a correct server, let $\client$ be a client, let $m \in \messages$.
If $\server$ eventually invokes $order\rp{\client, m, b}$ for any $b \in \altime$, then $\server$ eventually triggers $\event{\tobin}{Deliver}{\client, m}$.
\begin{proof}
By Observation \ref{observation:tob.deliveredgrows}, $delivered$ is initially empty at $\server$.
$\server$ adds $\rp{\client, m}$ to $delivered$ (line \ref{line:tob.adddelivered} only) only after triggering $\event{\tobin}{Deliver}{\client, m}$ (line \ref{line:tob.deliver}).
Upon invoking $order\rp{\client, m, b}$ (line \ref{line:tob.order}), $\server$ either satisfies $\rp{\client, m} \in delivered$ or $\rp{\client, m} \notin delivered$.
If $\rp{\client, m} \in delivered$, then $\server$ already triggered $\event{\tobin}{Deliver}{\client, m}$.
If $\rp{\client, m} \notin delivered$ (line \ref{line:tobc.deliverdecision}), then $\server$ triggers $\event{\tobin}{Deliver}{\client, m}$ (line \ref{line:tob.deliver}), and the lemma is proved.
\end{proof}
\end{lemma}

\begin{corollary}
\label{corollary:tob.filteredaredelivered}
Let $\rp{\client, m, b} \in \filtered$.
Every correct server eventually triggers $\event{\tobin}{Deliver}{\client, m}$.
\begin{proof}
It follows immediately from Lemma \ref{lemma:tob.filteredareordered}, Definition \ref{definition:tob.order} and Lemma \ref{lemma:tob.orderedaredelivered}.
\end{proof}
\end{corollary}

\begin{lemma}
\label{lemma:tob.submitinsequence}
Let $\client$ be a correct client, let $m \in \messages$, let $K \in \mathbb{N}$ such that $\client$ invokes $submit\rp{m, \any}$ at least $K$ times.
Let $k \leq K$.
The $k$-th time $\client$ invokes $submit\rp{m, \any}$, $\client$ invokes $submit\rp{m, k - 1}$.
\begin{proof}
We prove the lemma by induction on $k$.
Let assume $K \geq 1$.
Upon initialization, $\client$ sets $submissions = \cp{}$ (line \ref{line:tobc.initsubmissions}).
$\client$ sets $submission\qp{m}$ (line \ref{line:tobc.setsubmission} only) only upon executing $submit\rp{m, \any}$ (line \ref{line:tobc.submit}).
This means that, before ever invoking $submit\rp{m, \any}$, $\client$ satisfies $m \notin submissions$.
As a consequence, $\client$ cannot invoke $order\rp{m, \any}$ for the first time from line \ref{line:tobc.submituponfailure}, as that would require $m \in submissions$ (line \ref{line:tobc.failurecondition}).
The only other possibility is that $\client$ first invokes $order\rp{m, \any}$ from line \ref{line:tobc.submituponbroadcast}.
We underline that $\client$ does so only upon triggering $\event{\tobin}{Broadcast}{m}$ (this will be useful later in the proof).
Upon invoking $order\rp{m, \any}$ from line \ref{line:tobc.submituponbroadcast}, $\client$ invokes $order\rp{m, 0}$, proving the induction for $k = 1$.

Let us assume that the induction holds for some $k < K$: upon invoking $submit\rp{m, \any}$ for the $k$-th time, $\client$ invoked $submit\rp{m, k - 1}$.
Because $\client$ first invoked $submit\rp{m, \any}$ upon triggering $\event{\tobin}{Broadcast}{m}$, $\client$ cannot invoke $submit\rp{m, \any}$ for the $\rp{k + 1}$-th time from line \ref{line:tobc.submituponbroadcast}, as no correct client triggers $\event{\tobin}{Brodcast}{m}$ more than once.
The only other possibility is that $\client$ invokes $order\rp{m, \any}$ for the $\rp{k + 1}$-th time from line \ref{line:tobc.submituponfailure}.
By induction hypothesis, upon last invoking $submit\rp{m, \any}$ (line \ref{line:tobc.submit}), $\client$ set $submissions\qp{m} = \rp{k - 1, \any}$ (line \ref{line:tobc.setsubmission}).
Because $\client$ only sets $submissions\qp{m}$ upon executing $submit\rp{m, \any}$, when $\client$ invokes $submit\rp{m, \_}$ for the $\rp{k + 1}$-th time, $\client$ still satisfies $submissions\qp{m} = \rp{\rp{k - 1}, \any}$.
This proves (line \ref{line:tobc.submituponfailure}) that the $\rp{k + 1}$-th time $\client$ invokes $submit\rp{m, \any}$, $client$ invokes $submit\rp{m, k}$.
By induction, the lemma is proved.
\end{proof}
\end{lemma}

\begin{lemma}
\label{lemma:tob.submituntildeliver}
Let $\client$ be a correct client, let $m \in \messages$ such that $\client$ triggers $\event{\tobin}{Broadcast}{m}$.
If no correct server ever triggers $\event{\tobin}{Deliver}{\client, m}$, then $\client$ invokes $submit\rp{m, \any}$ an infinite number of times.
\begin{proof}
Let us assume that no correct server ever triggers $\event{\tobin}{Deliver}{\client, m}$. 
We prove by induction that, for all $k \in \mathbb{N}$, $\client$ invokes $submit\rp{m, \any}$ at least $k$ times.
Upon triggering $\event{\tobin}{Broadcast}{m}$ (line \ref{line:tobc.broadcast}), $\client$ invokes $submit\rp{m, \_}$ (line \ref{line:tobc.submituponbroadcast}). 
This proves the induction for $k = 1$.

Let us assume that induction holds for some $k \geq 1$.
Let $submit\rp{m, r}$ capture $\client$'s $k$-th $submit\rp{m, \any}$ invocation.
Upon executing $submit\rp{m, r}$ (line \ref{line:tobc.submit}), $\client$ sends the same $\qp{\texttt{Message, m, b}}$ to every server (lines \ref{line:tobc.sendmessagefor} and \ref{line:tobc.sendmessage}), for some $b \in \altime$ (line \ref{line:tobc.bet}).
$\client$ then sets $submissions\qp{m} = \rp{r, b}$ (line \ref{line:tobc.setsubmission}).
Noting that $\client$ updates $submissions\qp{m}$ (line \ref{line:tobc.setsubmission}) only upon executing $submit\rp{m, \any}$ (line \ref{line:tobc.submit}), $submission\qp{m}$ stays unchanged until $\client$ invokes $submit\rp{m, \any}$ for the $\rp{k + 1}$-th time.

Upon delivering $\client$'s $\qp{\texttt{Message}, m, b}$ message (line \ref{line:tob.delivermessage}), every correct server triggers $\event{\conin\qp{\client, m, b}}{Propose}{}$.
By the Termination property of $\conin\qp{\rp{\client, m, b}}$ and Corollary \ref{corollary:tob.filteredaredelivered}, every correct server eventually triggers $\event{\conin\qp{\rp{\client, m, b}}}{Decide}{\false}$.
Upon doing so (line \ref{line:tob.decide}) every correct server sends a $\qp{\texttt{Decision}, m, b, \false}$ back message to $\client$ (line \ref{line:tob.senddecision}).

Let $\server$ be a correct server.
Upon delivering a $\qp{\texttt{Decision}, m, b, \false}$ message from $\server$ (line \ref{line:tobc.deliverdecision}), $\client$ sets $decisions\qp{\rp{m, b, \server}} = \false$ (line \ref{line:tobc.setdecision}).
$\server$ updates $decisions\qp{\rp{m, b, \server}}$ to some $v \in \values$ (line \ref{line:tobc.setdecision} only) only upon delivering a $\qp{\texttt{Decision}, m, b, v}$ message from $\server$ (line \ref{line:tobc.deliverdecision}).
In turn, $\server$ sends a $\qp{\texttt{Decision}, m, b, v}$ (line \ref{line:tob.senddecision} only) only upon triggering $\event{\conin\qp{\rp{\client, m, b}}}{Decide}{v}$ (line \ref{line:tob.decide}).
By the No Duplication property of $\conin\qp{\rp{\client, m, b}}$, we then have that $\client$ never updates $decisions\qp{\rp{m, b, \server}}$ again.

As we assume at least $\faultcount + 1$ correct servers, eventually $\server$ satisfies
\begin{equation*}
    \abs{\server \such decisions\qp{\rp{m, b, \server}} = \false} \geq \faultcount + 1
\end{equation*}
Because $\client$ satisfies $submissions\qp{m} = \rp{r, b}$ until $\client$ invokes $submit\rp{m, \any}$ for the $\rp{k + 1}$-th time, $\client$ eventually triggers the condition at line \ref{line:tobc.failurecondition}, and invokes $submit\rp{m, \any}$ (line \ref{line:tobc.submituponfailure}).
This proves that the induction holds for $k + 1$ and concludes the lemma.
\end{proof}
\end{lemma}

\begin{corollary}
\label{corollary:tob.submitlongenough}
Let $\client$ be a correct client, let $m \in \messages$ such that $\client$ triggers $\event{\tobin}{Broadcast}{m}$.
If no correct server ever triggers $\event{\tobin}{Deliver}{m}$, then $\client$ eventually invokes $submit\rp{m, r}$ for every $r \in \mathbb{N}$.
\begin{proof}
It follows immediately from Lemmas \ref{lemma:tob.submitinsequence} and \ref{lemma:tob.submituntildeliver}.
\end{proof}
\end{corollary}

\begin{theorem}
\label{theorem:tob.validity}
\tobal satisfies Validity.
\begin{proof}
Let $\client$ be a correct client, let $m \in \messages$ such that $\client$ triggers $\event{\tobin}{Broadcast}{m}$.
Let us assume, for the sake of contradiction, that no correct server ever triggers $\event{\tobin}{Deliver}{\client, m}$.
By Corollary \ref{corollary:tob.submitlongenough}, $\client$ eventually invokes $submit\rp{m, r}$ for every $r \in \mathbb{N}$.

Let $r^* \in \mathbb{N}$ such that 
\begin{equation*}
    2^{r^*}  > \frac{\messagedelay + 2 \timedrift}{\hat \messagedelay}
\end{equation*}
At some time $t \in \altime$, $\client$ invokes $submit\rp{m, r^*}$ (line \ref{line:tobc.submit}).
As we assume partial synchrony, and noting that $\betmargin > 0$ (line \ref{line:tob.betmargin}), upon evaluating line \ref{line:tobc.bet} $\client$ sets $b > t - \timedrift + 2^{r^*}\hat \messagedelay$.
$\client$ then sends a $\qp{\texttt{Message}, m, b}$ message to every server (lines \ref{line:tobc.sendmessagefor} and \ref{line:tobc.sendmessage}). 

Let $\server$ be any correct server.
$\server$ delivers $\client$'s $\qp{\texttt{Message}, m, b}$ message by time $t + \messagedelay$.
When $\server$ does so (line \ref{line:tob.delivermessage}), $\server$'s local time is at most $t + \messagedelay + \timedrift$.
As a result, upon evaluating line \ref{line:tob.intime}, $\server$ finds
\begin{equation*}
    b > t - \timedrift + 2^{r^*}\hat\messagedelay > t + \messagedelay + \timedrift \geq local\_time\rp{}
\end{equation*}
and triggers $\event{\conin\qp{\rp{\client, m, b}}}{Propose}{\true}$ (line \ref{line:conal.propose}).
By the \validity property of $\conin\qp{\rp{\client, m, b}}$, every correct server eventually triggers $\event{\conin\qp{\rp{\client, m, b}}}{Decide}{\true}$.
By Corollary \ref{corollary:tob.filteredaredelivered}, every correct server eventually triggers $\event{\conin}{Deliver}{\client, m}$.
This contradicts no correct server ever triggering $\event{\conin}{Deliver}{\client, m}$ and proves the theorem.
\end{proof}
\end{theorem}

\subsection{Good-Case Latency}
\label{subsection:tob.latency}

All results in this section rest on the good-case assumptions introduced in \cref{subsection:background.goodcase}.
We also assume that all \con instances in $\conin$ are implemented by \conal.

\begin{lemma}
\label{lemma:tob.observedcamefrommessage}
Let $\rp{\client, m, b} \in \universe$. $\client$ eventually issues a $\qp{\texttt{Message}, m, b}$ message.
\begin{proof}
Let $\server$ denote the server that first adds $\rp{\client, m, b}$ to $observed$.
$\server$ does so (line \ref{line:tob.addobserved} only) only upon executing $spot\rp{\client, m, b}$ (line \ref{line:tob.spot}).
$\server$ invokes $spot\rp{\client, m, b}$ only from lines \ref{line:tob.spotuponmessage} and \ref{line:tob.spotuponobserve}.
If $\server$ invokes $spot\rp{\client, m, b}$ from line \ref{line:tob.spotuponmessage}, then $\server$ delivered a $\qp{\texttt{Message}, m, b}$ from $\client$ (line \ref{line:tob.delivermessage}) and the lemma trivially holds true.

Let us assume, for the sake of contradiction, that $\server$ invokes $spot\rp{\client, m, b}$ from line \ref{line:tob.spotuponobserve} instead.
$\server$ does so upon delivering an $\qp{\texttt{Observe}, \client, m, b}$ message from some server $\serversecondary$ (line \ref{line:tob.deliverobserve}).
Instantaneously after issuing an $\qp{\texttt{Observe}, \client, m, b}$ message (line \ref{line:tob.sendobserve} only) $\serversecondary$ adds $\rp{\client, m, b}$ to $observed$.
This contradicts $\server$ being the first server to add $\rp{\client, m, b}$ to observed, proves that $\server$ invokes $spot\rp{\client, m, b}$ from line \ref{line:tob.spotuponmessage}, and concludes the lemma.
\end{proof}
\end{lemma}

\begin{lemma}
\label{lemma:tob.allmessagesintime}
Let $\client$ be a client, let $m \in \messages$, let $b \in \altime$ such that $\client$ issues a $\qp{\texttt{Message}, m, b}$ message.
Every server delivers $\qp{\texttt{Message}, m, b}$ by time $b - \betmargin$.
\begin{proof}
$\client$ issues a $\qp{\texttt{Message}, m, b}$ message (line \ref{line:tobc.sendmessage} only) only upon executing $submit\rp{m, r}$ (line \ref{line:tobc.submit}) for some $r \geq 0$.
Let $t$ denote the time when $\client$ invokes $submit\rp{m, r}$.
Upon doing so (line \ref{line:tobc.submit}), $\client$ instantaneously evaluates
\begin{equation*}
    b = t + 2^r \hat \messagedelay + \betmargin = t + 2^r \messagedelay + \betmargin \geq t + \messagedelay + \betmargin
\end{equation*}
(line \ref{line:tobc.bet}) and sends a $\qp{\texttt{Message}, m, b}$ to every server (lines \ref{line:tobc.sendmessagefor} and \ref{line:tobc.sendmessage}).
As a result, every server delivers $\qp{\texttt{Message}, m, b}$ by time $t + \messagedelay = b - \betmargin$, which proves the lemma.
\end{proof}
\end{lemma}

\begin{lemma}
\label{lemma:tob.lockafterlocal}
Every server always satisfies $lock\_time\rp{} \leq local\_time\rp{}$.
\begin{proof}
Let $\server$ be a server, let $t \in \altime$.
Let us assume, for the sake of contradiction, that at time $t$ $\server$ satisfies $lock\_time\rp{} > \rp{local\_time\rp{} = t}$.
This means (line \ref{line:tob.locktimereturn}) that at least $4\faultcount + 1$ values in $remote\_times$ are greater than $t$.
Let $\serversecondary$ be any server such that, at time $t$, $\server$ satisfies $remote\_times\qp{\serversecondary} > t$.
$\server$ sets $remote\_times\qp{\serversecondary} > t$ (line \ref{line:tob.updateremotetimes} only) only upon delivering a $\qp{\texttt{Time}, t' > t}$ message from $\serversecondary$ (line \ref{line:tob.delivertime}).
This means that, by time $t$, $\serversecondary$ issues a $\qp{\texttt{Time}, t'}$ message.
Noting that $\serversecondary$ only issues a $\qp{\texttt{Time}, t'}$ message at time $t'$ (line \ref{line:tob.sendtime} only), however, $\serversecondary$ cannot issue a $\qp{\texttt{Time}, t'}$ message by $t < t'$.
We thus reach a contradiction, and the lemma is proved.
\end{proof}
\end{lemma}

\begin{lemma}
\label{lemma:tob.proposedmeansproposed}
Let $\server$ be a server. 
Let $\client$ be a client, let $m \in \messages$, let $b \in \altime$ such that $\server$ adds $\rp{\client, m, b}$ to $proposed$ before time $b$.
$\server$ triggers $\event{\conin\qp{\rp{\client, m, b}}}{Propose}{\true}$.
\begin{proof}
$\server$ adds $\rp{\client, m, b}$ to $proposed$ (lines \ref{line:tob.addproposeduponmessage} and \ref{line:tob.addproposeduponexpire} only) only after triggering $\event{\conin\qp{\rp{\client, m, b}}}{Propose}{}$ (lines \ref{line:tob.proposeuponmessage} and \ref{line:tob.proposeuponexpire}).
This means that $\server$ triggers $\event{\conin\qp{\rp{\client, m, b}}}{Propose}{}$ before time $b$.
$\server$, however, triggers $\event{\conin\qp{\rp{\client, m, b}}}{Propose}{\false}$ (lines \ref{line:tob.proposeuponmessage} and \ref{line:tob.proposeuponexpire} only) only if $local\_time\rp{} \geq b$ (lines \ref{line:tob.intime} and \ref{line:tob.proposeexpirecondition}).
This proves that $\server$ triggers $\event{\conin\qp{\rp{\client, m, b}}}{Propose}{\true}$ and concludes the lemma.
\end{proof}
\end{lemma}

\begin{lemma}
\label{lemma:tob.timelypropose}
Let $\rp{\client, m, b} \in \universe$.
By time $b - \betmargin$, every server adds $\rp{\client, m, b}$ to $candidates$ and triggers $\event{\conin\qp{\rp{\client, m, b}}}{Propose}{\true}$.
\begin{proof}
Let $\server$ be a server.
By Lemmas \ref{lemma:tob.observedcamefrommessage} and \ref{lemma:tob.allmessagesintime}, $\server$ delivers a $\qp{\texttt{Message}, m, b}$ message from $\client$ at some time $t \leq b - \betmargin$.
Upon doing so (line \ref{line:tob.delivermessage}), $\server$ instantaneously invokes $spot\rp{\client, m, b}$ (line \ref{line:tob.spotuponmessage}, observes that
\begin{equation*}
    b \geq t + \betmargin > local\_time\rp{} \geq lock\_time\rp{}
\end{equation*}
(line \ref{line:tob.candidatecondition}) and adds $\rp{\client, m, b}$ to $candidates$ (line \ref{line:tob.addcandidate}).

Upon returning from $spot\rp{\client, m, b}$, $\serversecondary$ instantaneously moves on to check if $\rp{\client, m, b} \in proposed$.
If $\rp{\client, m, b} \in proposed$, then by Lemma \ref{lemma:tob.proposedmeansproposed} $\serversecondary$ triggered $\event{\conin\qp{\rp{\client, m, b}}}{Propose}{\true}$.
If $\rp{\client, m, b} \notin proposed$ (line \ref{line:tob.proposedcheckuponmessage}), then $\serversecondary$ verifies
\begin{equation*}
    b \geq t + \betmargin > local\_time
\end{equation*}
(line \ref{line:tob.intime} and instantaneously triggers $\event{\conin\qp{\rp{\client, m, b}}}{Propose}{\true}$ (line \ref{line:tob.proposeuponmessage}).
This proves that $\server$ triggers $\event{\conin\qp{\rp{\client, m, b}}}{Propose}{\true}$ by time $b - \betmargin$ and concludes the lemma.
\end{proof}
\end{lemma}

\begin{corollary}
\label{corollary:tob.timelydecide}
Let $\rp{\client, m, b} \in \universe$.
By time $b + \messagedelay - \betmargin$, every server triggers $\event{\conin\qp{\rp{\client, m, b}}}{Decide}{\true}$.
\begin{proof}
It follows immediately from Lemma \ref{lemma:tob.timelypropose} and Theorem \ref{theorem:consensus.latency}.
\end{proof}
\end{corollary}

\begin{lemma}
\label{lemma:tob.timelylock}
Let $\rp{\client, m, b} \in \universe$.
By time $b + \messagedelay$, every server satisfies $lock\_time\rp{} \geq b$.
\begin{proof}
We start by noting that, upon adding $\rp{\client, m, b}$ to $observed$ (line \ref{line:tob.addobserved} only), a correct server schedules $beat\rp{}$ for execution at time $b$ (line \ref{line:tob.schedulebeat}).

Let $\serversecondary$ be a server.
By Lemmas \ref{lemma:tob.observedcamefrommessage} and \ref{lemma:tob.allmessagesintime}, $\serversecondary$ delivers a $\qp{\texttt{Message}, m, b}$ message from $\client$ by time $b - \betmargin$.
Upon doing so (line \ref{line:tob.delivermessage}), $\serversecondary$ instantaneously invokes $spot\rp{\client, m, b}$ (line \ref{line:tob.spotuponmessage}).
By Lemma \ref{lemma:tob.spot}, $\serversecondary$ adds $\rp{\client, m, b}$ to $observed$ before time $b$.
As a result, $\serversecondary$ executes $beat\rp{}$ at time $b$.
Upon doing so (line \ref{line:tob.beat}), $\serversecondary$ sends a $\qp{\texttt{Time}, b}$ message to every server (lines \ref{line:tobc.sendmessagefor} and \ref{line:tobc.sendmessage}).
In summary, at time $b$, every server sends a $\qp{\texttt{Time}, b}$ message to every server.

Let $\server$ be a server.
By time $b + \messagedelay$, $\server$ delivers a $\qp{\texttt{Time}, b}$ message from every server.
Upon doing so (line \ref{line:tob.delivertime}), $\server$ instantaneously sets every value in $remote\_times$ to a value greater or equal to $b$ (line \ref{line:tob.updateremotetimes}).
Having done so, by Observation \ref{observation:remotetimesgrow}, $\server$ forever satisfies $remote\_times\qp{\serversecondary} \geq b$ for every server $\serversecondary$.
Because $lock\_time\rp{}$ (line \ref{line:tob.locktime}) returns a value greater or equal to $b$ if at least $4 \faultcount + 1$ values in $remote\_times$ are greater or equal to $b$ (line \ref{line:tob.locktimereturn}), $\server$ forever satisfies $lock\_time\rp{} \geq b$, and the lemma is proved.
\end{proof}
\end{lemma}

\begin{lemma}
\label{lemma:tob.timelydelivery}
Let $\rp{\client, m, b} \in \universe$.
By time $b + \messagedelay$, every server triggers $\event{\tobin}{Deliver}{\client, m}$.
\begin{proof}
We start by noting that, by Lemma \ref{lemma:tob.timelypropose} and Corollary \ref{corollary:tob.timelydecide}, we have
\begin{equation*}
    \universe = \candidates{\server} = \filtered
\end{equation*}
Moreover, again by Lemma \ref{lemma:tob.timelypropose} and Corollary \ref{corollary:tob.timelydecide}, by time $b + \messagedelay$, $\server$ satisfies 
\begin{align*}
    \candidates{\server}[\leq b] &\subseteq candidates \\
    \candidates{\server}[\leq b] &\subseteq decisions
\end{align*}
Finally, by Lemma \ref{lemma:tob.timelylock}, by time $b + \messagedelay$, $\server$ satisfies
\begin{equation*}
    lock\_time\rp{} \geq b
\end{equation*}
As such, until $last\_processed \geq \rp{\client, m, b}$, $\server$ keeps satisfying the condition at line \ref{line:tob.processcondition}.

Unless $last\_processed = \rp{\bot, \bot, -\infty}$ (line \ref{line:tob.initlastprocessed}), we have $last\_processed \in \candidates{\server}$.
Indeed, $\server$ sets $last\_processed = \rp{\client', m', b'}$ (line \ref{line:tob.updatelastprocessed}) only if $\rp{\client', m', b'} \in candidates$ (line \ref{line:tob.processcondition}). 
Moreover, by Observation \ref{observation:tob.lastprocessedgrows}, $last\_processed$ is non-decreasing. 
Finally, because by Lemma \ref{lemma:tob.croppedcandidatesarefinite} we have $\abs{\candidates{\server}[\leq b]} < \infty$, $\server$ can update the value of $last\_processed$ only a finite number of times before $last\_processed \geq \rp{\client, m, b}$. 

To summarize, by time $b + \messagedelay$, $\server$ keeps instantaneously triggering the condition at line \ref{line:tob.processcondition} until $last\_processed \geq \rp{\client, m, b}$.
Every time $\server$ triggers the condition at line \ref{line:tob.processcondition}, $\server$ updates $last\_processed$ (line \ref{line:tob.updatelastprocessed}) to a larger element of $\candidates{\server}$ (line \ref{line:tob.processcondition}).
Having instantaneously done so a finite number of times, $\server$ satisfies $last\_processed \geq \rp{\client, m, b}$.
As such, by Lemma \ref{lemma:tob.lastprocessedovertake}, $\server$ instantaneously invokes $order\rp{\client, m, b}$.
By Lemma \ref{lemma:tob.orderedaredelivered}, $\server$ also instantaneously triggers $\event{\tobin}{Deliver}{\client, m}$, and the lemma is proved.
\end{proof}
\end{lemma}

\begin{theorem}
\label{theorem:tob.latency}
Let $\client$ be a client, let $m \in \messages$, let $t \in \altime$ such that $\client$ triggers $\event{\tobin}{Broadcast}{m}$ at time $t$.
Every server triggers $\event{\tobin}{Deliver}{\client, m}$ by time $t + 2\messagedelay + \betmargin$.
\begin{proof}
Upon triggering $\event{\tobin}{Broadcast}{m}$ (line \ref{line:tobc.broadcast}), $\client$ instantaneously invokes $submit\rp{m, 0}$ (line \ref{line:tobc.submituponbroadcast}), thus sending a $\qp{\texttt{Message}, m, b}$ message to every server, with
\begin{equation*}
    b = t + \hat \messagedelay + \betmargin = t + \messagedelay + \betmargin
\end{equation*}
(lines \ref{line:tobc.sendmessagefor} and \ref{line:tobc.sendmessage}).
Upon delivering $\server$'s $\qp{\texttt{Message}, m, b}$ message (line \ref{line:tob.delivermessage}), a server invokes $spot\rp{\client, m, b}$ (line \ref{line:tob.spotuponmessage}).
By Lemma \ref{lemma:tob.spot} and Definition \ref{definition:tob.universe} we then have $\rp{\client, m, b} \in \universe$, and the theorem follows from Lemma \ref{lemma:tob.timelydelivery}.
\end{proof}
\end{theorem}

\subsection{Quasi-Optimality of the Good-Case Latency}
\label{appendix:tob.quasioptimal}

In this section, we prove that \tobal's good-case latency is \emph{quasi-optimal}, meaning that no \tob implementation can undercut \tobal's good-case latency by any finite amount.
As we proved in \cref{subsection:tob.latency}, in the good case (see \cref{subsection:background.goodcase}) \tobal attains a broadcast-to-delivery latency of $2 \messagedelay + \epsilon$, where $\messagedelay$ represents the network's message delay and $\epsilon$ is an arbitrarily small constant (\eg the smallest unit of time that a computer can represent).
Here, we prove \tobal's quasi-optimality by showing that no \tob implementation can achieve a latency smaller than $2 \messagedelay$, not even in the good case.
Whether or not a good-case latency of \emph{exactly} $2 \messagedelay$ is achievable remains an open problem, albeit one with limited practical implications, as \tobal's good-case latency is already arbitrarily close to $2 \messagedelay$.

The impossibility of \tob in less than two message delays is fairly intuitive. 
Let us consider the example of a client broadcasting some message $m$ at some time $t$.
Let $\server$ be a correct server.
In order to deliver $m$ before $t + 2 \messagedelay$, $\server$ would have to deliver $m$ based only on the messages it directly receives from $\client$, thus foregoing any form of coordination with the other servers.
Indeed, two message delays are required for $\server$ to simply determine whether or not the other servers received $m$ from $\client$: one for $\client$ to disseminate $m$ to the other servers, and one for the other servers to report their observations to $\server$.

\begin{definition}[Bounded-latency \tob implementation]
\label{definition:timeboundedtob}
Let $\algo$ be an implementation of \tob under the assumptions presented in \cref{subsection:background.model}.
$\algo$ is \textbf{$T$-bounded} if and only if, for every execution $\exec$ of $\algo$ abiding by the good-case assumptions presented in \cref{subsection:background.goodcase}, if a client $\client$ broadcasts a message $m$ by time $t$, then every server delivers $m$ by time $t + T$.
\end{definition}

We underline that, by \cref{definition:timeboundedtob} and \cref{theorem:tob.latency}, Flutter is bounded by $2 \messagedelay + \epsilon$ for every $\epsilon > 0$.

\begin{theorem}
Let $T < 2 \messagedelay$. No \tob implementation is $T$-bounded.
\begin{proof}
Let us assume towards a contradiction (see \cref{definition:timeboundedtob}) that some $T$-bounded implementation $\algo$ of \tob does exists. We can reach a contradiction using indistinguishable executions.

Let $\exec$, $\exec'$ be executions of $\algo$, abiding by the good-case assumptions presented in \cref{subsection:background.goodcase}, such that: the delay of every message in $\exec$ / $\exec'$ is exactly $\messagedelay$; throughout all $\exec$ (resp., $\exec'$), a single client $\client$ broadcasts a single message $m$ (resp., $m' \neq m$) at some time $t_0$.
Let $\server$, $\server' \neq \server$ be servers.
By \cref{definition:timeboundedtob}, in $\exec$ (resp., $\exec'$), $\server$ (resp., $\server'$) delivers $m$ (resp., $m'$) as first and only message (see \cref{subsection:background.abstractions}) at some time $t$ (resp., $t'$), with $t, t' \leq t_0 + T$.

We start by noting that $\exec$ and $\exec'$ are identical up to $t_0$.
Indeed, prior to $t_0$, no process is provided with any input in either execution, and $\algo$ unfolds deterministically: the processes issue the same messages at the same times, each message is delivered with the same delay, resulting in identical executions for $\exec$ and $\exec'$.
Moreover, between $t_0$ and $t_0 + \messagedelay$, the execution of every process other than $\client$ is identical in $\exec$ and $\exec'$.
Indeed, before $t_0 + \messagedelay$, a process receives only messages that were sent before $t_0$.
By the above, this means that every process other than $\client$ receives the same messages at the same times, unfolding deterministically and identically in both $\exec$ and $\exec'$.
Finally, let $\chi$ be any process other than $\client$.
Before $t_0 + 2 \messagedelay$, $\server$ (resp., $\server'$) receives the same sequence of messages from $\chi$ in both $\exec$ and $\exec'$.
Indeed, before $t_0 + 2\messagedelay$, $\server$ (resp., $\server'$) receives only messages that were sent before $t_0 + \messagedelay$, and by the above, before $t_0 + \messagedelay$, $\chi$ issues the same sequence of messages in both $\exec$ and $\exec'$.

Now, let $\exec^*$ be a variant of $\exec$ / $\exec'$, identical in all aspects except that $\client$ is faulty.
In $\exec^*$, $\client$ sends to $\server$ the same messages it would send in $\exec$, and to $\server'$ the same messages it would send in $\exec'$. 
By the above, $\server$ receives the same messages at the same times in both $\exec$ and $\exec*$, thus delivering $m$ at time $t \leq t_0 + T < t_0 + 2 \messagedelay$.
Similarly, $\server'$ delivers $m'$ at time $t'$. 
In summary, in $\exec^*$ $\server$ and $\server'$ respectively deliver $m$ and $m' \neq m$ as first and only message.
This violates the Agreement and Total Order property (see \cref{subsection:background.abstractions}) and contradicts $\algo$ being an implementation of \tob, thus concluding the theorem.
\end{proof}
\end{theorem}

\subsection{Denial of Service}
\label{subsection:tob.dos}

\tobal spawns an instance of \con for every message and bet that a client submits to \tob.
Such an amplification makes \tobal vulnerable to denial-of-service attacks, wherein a malicious client overwhelms the system with spurious or useless requests.
While a full discussion of these attacks is beyond the scope of this theoretical paper, in this section we briefly argue that \tobal's vulnerability to denial of service is not qualitatively different from that of other \tob solutions, and that the same mitigation strategies that apply to those also apply to \tobal.

\paragraph{One consensus per broadcast.}
It is a fundamental result that \tob is as powerful a primitive as Consensus~\cite{ct96,dds87}.
Indeed, implementing Consensus using \tob is fairly straightforward: each process broadcast its proposal; upon delivering $4 \faultcount + 1$ proposals, a process decides the majority proposal, resorting to a deterministic tiebreak in the case of a draw.
As such, we believe that no \tob implementation can deliver a message without solving a problem equivalent to Consensus.

\paragraph{Batching.}
\tob implementations in systems literature employ a variety of batching techniques to amortize the cost of Consensus~\cite{pbft-tocs02,bft-fire-nsdi08,narwhal-tusk-eurosys22,chopchop-osdi24,autobahn-sosp24}.
Broadly speaking, batching involves multiple clients entrusting their messages to the same \emph{broker}, which packages them for bulk delivery by the servers.
Batched solutions usually offer a trade-off between latency and efficiency, as brokers wait to collect multiple messages in the same batch.
At a similar cost in terms of latency, \tobal can also accommodate batching.
A batched variant of \tobal would shift to brokers the responsibility of selecting a batch's bet; clients would authenticate their messages with signatures or symmetric message authentication codes.

\paragraph{Broadcast vs. broadcast attempt.}
A noteworthy design difference between \tobal and other \tob implementations is that \tobal spawns an instance of \con every time a client \emph{attempts} to broadcast a message.
Upon failing to get its message delivered, a client updates its bet and tries again, thus triggering a new instance of \con.
Such a difference, however, does not qualitatively affect \tobal's vulnerability to denial of service.
Even against a system that guarantees one delivery per Consensus instance, a malicious client could simply broadcast messages that are malformed, invalid or semantically ``useless''.
In doing so, the client would still force the expenditure of a Consensus instance for every message it disseminates, while also pulling the application layer into the computation required to process or discard the message.

\paragraph{Client authentication and rate limiting.}
As we discussed in \cref{subsection:background.model}, \tobal assumes a permissionless set of clients.
In settings where denial-of-service attacks are an issue, such an assumption can be dangerous.
It is worth underlying, however, that \tobal trivially generalizes to the permissioned setting.
As links are authenticated, servers can easily ignore messages from unauthorized clients~\cite{consensus-dos-resilience-asiaccs24}.
Similarly, \tobal is compatible with rate-limiting~\cite{consensus-dos-resilience-asiaccs24} (\eg each client gets to broadcast once per minute).
Such systems-oriented mechanisms, however, are beyond the scope of this theoretical paper.

\paragraph{Exponential backoff.}
Correct \tobal clients double their bet margin every time their messages are rejected.
This behavior can also be forced onto malicious clients.
To mitigate denial-of-service attacks, servers could force a client that failed to broadcast $n$ times to issue bets spaced by at least $2^n$ units of time.
Such a simple technique quickly neutralizes any mis-betting client.

\paragraph{Leader denial-of-service.}
Finally, it is worth underlining that \tobal's leaderless nature naturally sidesteps the scenario where an attacker overwhelms with requests whichever server is currently the leader, repeatedly causing expensive suspicion and re-election.

\end{document}